\documentclass{article}
\usepackage[utf8]{inputenc}
\usepackage{placeins}
\usepackage{bm}
\usepackage{bbm}
\usepackage{amsmath}
\usepackage{mathtools}
\DeclarePairedDelimiter\floor{\lfloor}{\rfloor}
\usepackage{multirow}
\usepackage{amssymb}
\usepackage{comment}
\usepackage{amsthm}
\usepackage{geometry}
\usepackage{graphicx}
\usepackage{setspace}
\usepackage{lineno}
\theoremstyle{definition}
\newtheorem{theorem}{Theorem}[section]

\newtheorem{lemma}{Lemma}[section]
\newtheorem{proposition}{Proposition}[section]
\newtheorem{assumption}{Assumption}
\newtheorem{definition}{Definition}

\title{Sequential Change-point Detection for Binomial Time Series}
\author{%
  Yajun Liu \footnote{Amazon.com, Seattle, WA, USA. Email: yajunliustacy@gmail.com}%
  \and Beth Andrews \footnote{Department of Statistics and Data Science, Northwestern University, Evanston, IL, USA. Email: bandrews@northwestern.edu}%
  }
\date{}

\begin{document}

\maketitle

\begin{abstract}
A binomial time series describes binary behaviors of individuals within a group, which depend on group behaviors in the past. Binomial time series data is widely applied in fields such as infection tracking and behavior analysis. In this paper, we introduce a generalized Binomial AR($p$) model with exogenous variables based on Generalized Linear Model (GLM), prove the statistical properties of the model when $p = 1$, and provide a parameter estimation method. Then, we propose a sequential change-point detection method for the generalized Binomial AR(1) model, facilitate real-time data monitoring and triggering alarms when a change point is detected. We apply the generalized Binomial AR(1) model to weekly pneumonia \& influenza mortality data and successfully identify change points related to the COVID-19 outbreak using the proposed method.
\end{abstract}

\section{Introduction}
In the time series literature, when we handle real-life data sets, we use appropriate models to fit the data and use the fitted models for further statistical inference and forecasting. The validity of forecasting relies on the assumption that the fitted models are still reasonable as more and more data are observed. To verify the assumption, we need to test if %the estimated parameters of 
the model remains the same. If there exist any parameters of the model changing at time $t$, we call $t$ a change point. Considering different circumstances, there are two main types of change-point detection: retrospective change-point detection and sequential change-point detection. Given a complete time series, retrospective change-point detection can be applied to find all existing change points in the data set. By matching the detected change points with historical events, we can gain insights into the relationships between the events and the time series of interest. Conversely, sequential change-point detection is used when we continue to observe a time series and would like to identify potential change points as they occur. Sequential change-point detection is important in practice, e.g. in data monitoring, risk control, etc.. The sooner we detect a change point, the sooner we can take action to control the risks associated with the change point and adjust the model.

In this work, we consider sequential change-point detection for Binomial AR processes, a topic which has not yet been fully developed. Among discrete-valued processes, Binomial AR models have a wide range of applications. For instance, customer engagement can be monitored by measuring the number of customers logging in or making purchases per unit of time among the $n$ customers sampled. Also, we can gain insights into the situation of a financial market by picking $n$ independent stocks that well represent the market and recording the number of stocks with increasing prices per unit time. As the data are observed, our goal is to detect a change in the underlying process as soon as possible so that we can prepare for new strategies. In practice, in order to monitor these kinds of data, developing sequential change-point detection on Binomial AR models is necessary. In this paper, we focus on sequential change-point detection for the Binomial AR(1) model based on generalized linear model. In brief, it depicts a binomial distribution with fixed and known total $n$ and time-varying probability that depends on the values of the previous output and exogenous variables at current time. The model can be used to describe behavior of a group that is dependent across time, e.g. customer engagement behavior of a website, the trends of stocks in a market, and so on. 

In the rest of the paper, we investigate the properties of the binomial time series and propose a sequential change-point detection method. In Section \ref{sec: Literature Review}, we go through some commonly used methods for change-point detection. A detailed introduction to change-point detection for binomial time series and the motivation for sequential change-point detection for Binomial AR observations are also provided. In Section \ref{sec: Model definition}, we first introduce the model. Then, we prove the properties of the process, cover the parameter estimation procedure and prove the consistency and the asymptotic normality of the estimated parameters when there is no change point. In Section \ref{sec: monitoring scheme}, we first introduce the structure of sequential change-point detection and the test statistic. Later, asymptotic distribution of the test statistic under the no change point null hypothesis is given and proved in Section \ref{sec: null}. Power analysis of the test statistic is discussed in Section \ref{sec: power analysis}. In Section \ref{sec: simulation}, we conduct several simulation studies to support the theoretical results in Section \ref{sec: null} and Section \ref{sec: power analysis}. In Section \ref{sec: real life data}, a real-life data analysis is given to demonstrate the use of the model.
% However, retrospective change-point detection no longer works when the no change point test needs to be conducted sequentially, as more and more data points appear. Therefore, sequential change-point detection was developed to fill the need of online detection.

% Given a whole data set, retrospective change-point detection can be applied to find all existing change points in the data set. By matching the detected change points with historical events, we can gain insights into the relationships between the events and the time series of interest. 
\section{Literature Review}
\label{sec: Literature Review}
Research on change-point detection originates from two papers written by Page \cite{page1954continuous}, \cite{page1955test}. The author raises the change-point detection question and proposes a detection procedure: monitoring the fluctuation of a cumulative score $S_n$, which represents the fluctuation of the observed data points, and taking action once $S_n$ exceeds a predetermined threshold. After these two papers, lots of research about retrospective change-point detection followed. Nowadays, theories about retrospective change-point detection for various time series models have been well developed. In general, researchers proposed various methods that can segment the whole data set so that the separated blocks have the largest differences. Following the cumulative score idea mentioned above, cumulative sum (CUSUM) type methods have been widely developed in various types of change-point detection problems. The basic idea of CUSUM is that assuming there is no change point in a data set, then after we segment the data set, the scaled cumulative sum of the observed data points (or other statistics that can represent the properties of the process) before the segmentation and after the segmentation should be close. In \cite{bai1994weak}, Bai develops a CUSUM test to detect parameter changes of an ARMA model. General CUSUM test for parameter change of stationary processes is then investigated in \cite{lee2003cusum}. For multivariate time series, Aue et al. \cite{aue2009break} use CUSUM to detect breaks in the covariance structure. Cho and Fryzlewicz \cite{cho2015multiple} propose using the threshold CUSUM statistic, which records a cumulative sum only when it exceeds a threshold, to conduct multiple-change-point detection. Besides various types of CUSUM, Kullback-Leibler information discrimination and likelihood ratio-based methods, which are also the methods focusing on measuring the difference %of statistics
of the process before and after a change point, are used in retrospective change-point detection, see \cite{last2008detecting} and \cite{davis1995testing}. Considering other methods, Davis et al.
% , Lee and Rodriguez-Yam 
apply minimum description length (MDL) to detect change points in piecewise autoregressive processes \cite{davis2006structural} and state-space models \cite{davis2008break}. MDL is capable of segmenting the whole data set in terms of the locations of the change points it detects and estimating the parameters of each segment by minimizing the code length describing the whole data set, which is defined as a sum of a measurement of the complexity of the models and a measurement of the portions of the data points that are not explained by the models. Ombao et al., in \cite{ombao2001automatic} and \cite{ombao2005slex}, use the smooth localized complex exponentials (SLEX) 
transform, which is closely related to Fourier transform, to convert and divide the whole data set so as to detect change points in bivariate piecewise stationary time series and multivariate piecewise stationary time series, respectively. The algorithm segments the time series into different blocks such that the cost function computed based on the SLEX transformation, which is a family of orthogonal basis functions, is minimized. More methods can be found in the review papers \cite{aue2013structural} and \cite{horvath2014extensions}.

Unlike retrospective change-point detection, sequential change-point detection
focuses more on methods that people can use to detect changes online. It is desired that we can detect the change point as soon as possible after it occurs in order to provide accurate forecasting and be alert to the reasons for the change point. Under this circumstance, the methodologies developed for retrospective change-point detection don't work anymore. Since retrospective change-point detection is based on the whole data set, false positive alarm will be eventually triggered with probability 1 as the test is conducted repeatedly every single time a new data point appears. In sequential change-point detection, we create a monitoring statistic that is updated once a new data point appears. Under the no change point null hypothesis, one commonly used idea is to derive the limiting behavior of the monitoring statistic, which can be a stochastic process (e.g. a Wiener process). That is, instead of conducting tests repeatedly on the whole data set, which is actually updated every time a new point appears, we create stochastic processes and investigate the dynamic behaviors of the processes in sequential change-point detection. Within the monitoring range, given a significance level $\alpha$, we are able to calculate the threshold that the stochastic process exceeds with probability $\alpha$, if there is no change point. By doing so, we set up a universal threshold for the monitoring statistic within the monitoring range and it guarantees that the overall Type I error of the test is $\alpha$. Because of that, the aforementioned false positive issue can be solved. The paper written by Chu et al. \cite{chu1996monitoring} is the origin of sequential change detection. The authors point out that the existing change-point detection algorithms can only be used to detect change points in historical data sets of fixed size, and give two sequential monitoring procedures and the corresponding asymptotically controlled sizes. After the paper by Chu et al., sequential detection for continuous models developed gradually. For detecting changes in the parameters of a linear regression model, Horv{\'a}th et al. derive asymptotic distributions for CUSUM type statistics based on residuals and recursive residuals in \cite{horvath2004monitoring}. Then, in \cite{horvath2007sequential}, the Darling-Erd{\H{o}}s theorem is established for sequential change detection in linear regression model. It's a commonly used tool for developing the limiting distributions, like Gumbel distribution and Laplace distribution, for the extreme values of test statistics, of which the asymptotic distributions are, in general, based on stochastic processes. With fewer restrictions on the error term, in \cite{aue2006change}, the CUSUM test is developed for linear regression when the error series is a martingale difference sequence. For the AR$(p)$ model, sequential change-point detection is developed based on the score vector, see \cite{gombay2009monitoring}. For GARCH$(p,q)$ model, quasi-likelihood score is used in sequential change-point detection in \cite{berkes2004sequential}. According to the common structure for sequential detection, the monitoring procedure would be stopped if the test statistic exceeds a predetermined threshold. Therefore, the distribution of the stopping time has also been investigated. For example, the asymptotic distribution of the stopping rule statistic, which is based on a CUSUM test, for detecting changes in the mean of the observations is obtained in \cite{aue2004delay} and \cite{aue2008monitoring}. The review paper \cite{polunchenko2012state} provides a summary of some major formulations for sequential change-point detection.

More recently, sequential change-point detection for various discrete-valued time series models has been gradually developed. In \cite{gut2002truncated} and \cite{gut2009truncated}, Gut and Steinebach propose the Darling-Erd{\H{o}}s theorems for different stopping rules in sequential change-point detection for renewal counting processes, which count the minimum number of i.i.d observations of which the cumulative sum exceeds a threshold. For Poisson AR (PAR) model, the asymptotic distribution of a test statistic based on the maximum likelihood estimator is derived in \cite{kengne2015sequential}, and a real data example: the number of transactions per minute for a stock, is provided to demonstrate the effectiveness of the method. Asymptotic results for sequential change-point detection of integer-valued autoregressive model of order 1 (INAR(1)) and integer-valued autoregressive conditional heteroskedasticity model of order 1 (INARCH(1)) are developed in \cite{hudecova2015detection}, \cite{hudecova2016change} and \cite{hudecova2017tests}, by constructing test statistics based on the corresponding empirical probability generating functions. The aforementioned models are for time series of counts when the current distribution is influenced by past counts and there is no constraint for the upper bound of the counts. Therefore, they can not depict behavior of specific individual(s) in a group when the total is fixed.

Time series with binary observation(s) from a specific individual(s) is another important branch of discrete-valued time series. For time series of a single observation, the retrospective change-point detection for binary autoregressive model of order $p$ (BAR$(p)$) is investigated in \cite{hudecova2013structural}, in which the Darling-Erd{\H{o}} theorem with respect to the normalized sum of residuals is derived. Similar results for BAR models can be found in \cite{kirch2015detection}, in which Kirch and Kamgaing propose the Darling-Erd{\H{o}} limit theorem based on the normalized score test statistic. What's more, 
an asymptotic distribution of the test statistic based on the Wiener process is given in the paper. Considering sequential change-point detection for discrete-valued time series, in \cite{kirch2015use}, Kirch and Kamgaing give certain regularity conditions under which the asymptotic results under the no change point null hypothesis and the behavior under alternatives hold. Then, they show that BAR(1) model satisfies the regularity conditions.

Extending from binary time series to binomial cases with fixed total $n>1$, time series of the number of independent successes within a group is also one of the important types of discrete-valued time series. Binomial time series models can be used when individual behavior that is revealed by a binary action (e.g. support/not support, increase/not increase, etc.) in a group is dependent on past observations. For example, customer engagement of a group of people of a website, which is measured by the number of customers logging in the website on a daily basis, is influenced by their past activity levels and the overall popularity of the website, and they can be reflected by the past login data of the same group of people. By modeling the process with a binomial time series model and conducting sequential change-point detection, we can detect change points, which represent alterations of customer engagement, as soon as possible so as to implement up-to-date marketing strategies. What's more, binomial time series is common in epidemiology and clinical trials, when the behavior of a group of people is the research interest. The same idea can be used in modeling financial trends of a group of individuals. Therefore, it's necessary to build binomial time series models, determine the properties of the processes and develop sequential change-point detection for the models. 

One type of model to depict binomial time series is binomial thinning process. McKenzie, in \cite{mckenzie1985some}, proposes the binomial thinning process of order 1 (denoted as Binomial AR(1) in the paper) that the number of successes at $t$ is dependent on two binomial distributions with different and fixed probabilities and totals equal to the number of successes at $t-1$ and the number of failures at $t-1$, respectively. Later, many papers about binomial thinning process follow. For example, \cite{weiss2009new} extends the Binomial AR(1) model to Binomial AR$(p)$, which applies the thinning process to the previous $p$ outputs. In \cite{scotto2014bivariate}, Scotto et al. make a bivariate extension of the Binomial AR(1) model and propose methods for parameter estimation and forecasting based on the bivariate binomial model. For change-point detection for the binomial thinning process, in \cite{yu2013change}, Yu et al. use likelihood ratio-based statistic to conduct sequential change-point detection and apply the method to influenza activity monitoring. 

However, binomial time series models based on generalized linear model (GLM) and change-point detection for the models haven't been developed yet. Binomial time series based on generalized linear model (GLM) follows a binomial distribution with varying probability dependent on past observations and exogenous variables through the link function. Compared to binomial thinning process, binomial time series based on generalized linear model (GLM) can include exogenous information and outputs earlier than $t-1$, which make the model more flexible and can explain a wider range of dependence. Meanwhile, the parameters in the link function allow us to interpret the influence of past observations and exogenous information on the current state explicitly and separately. So in this paper, we build a Binomial AR model based on GLM, and discuss the properties of the process and the estimated parameters when model order $p=1$. For the sake of simplicity, in the rest of the paper, we use Binomial AR($p$) to represent the Binomial AR($p$) model based on GLM. Following the method in \cite{kirch2015use}, sequential change-point detection for this model is also developed. 

%-------------------------------------------------------------------------------------------------------
\section{Binomial autoregressive model based on GLM}
\label{sec: Model definition}
In practice, many individual behaviors are influenced by the situation of the group they belong to. For example, the customer purchase behavior of a website at the current time depends on the overall situation of the economy, the popularity of the website, some events people are going through that can affect their purchase behaviors, and so on. These elements can be reflected by the purchase behaviors in the past several units of time. If we keep track of the buying behaviors of $n$ customers and record the number of customers among the $n$ people buying things on the website on a daily basis, then we should be able to reasonably predict the number of customers making purchases in the upcoming days if the elements affecting customer behavior do not change. Following this idea, we define the binomial time series model based on GLM.
\begin{equation}
\label{equation: model definition1}
    X_t|X_{t-1},X_{t-2},...,\mathbbm{Z}_{t-1},\mathbbm{Z}_{t-2},... \sim \text{Bin}(n,\pi_t(\bm{\beta})), \text{ with } g(n\pi_t(\bm{\beta})) :=g(\mu_t) =  \bm{\beta}^T \mathbbm{Z}_{t-1},
\end{equation}
in which $g(x) =\log(x/(n-x))$, and $\mathbbm{Z}_{t-1}$ is a regressor, which can include autoregressive variables, e.g. $\{X_{t-1},X_{t-2},...\}$ or other forms of past information, exogenous variables that affect the time-dependent probability $\pi_t(\bm{\beta})$, or a mixture of both. This model is inspired by the Binomial Logistic GARMA$(p, q)$ model proposed in \cite{benjamin2003generalized}. However, we remove the thresholds for $\{X_t\}$ and allow the output to be exact 0 or $n$. 
%%%%%%%%%%%%%%%%%%%%%%%%%%%% 
In this paper, we focus on the properties and sequential change-point detection for Binomial AR(1) model with exogenous variables $\{\bm{W}_t\}$ so that
\begin{itemize}
    \item they are i.i.d. random variables with known distribution function,
    \item $\{\bm{W}_t\}\in [a,b]^l$ with known $a$ and $b$,
    \item $\bm{W}_t$ is independent with $X_{t-i}, i = 1,2,...$.
\end{itemize}
That is,
\begin{equation}
\label{equation: model definition2}
\centering
\mathbbm{Z}_{t-1} = (1,X_{t-1},\bm{W}_t)^T, \text{ and } \bm{\beta} = (\varphi_0,\varphi_1,\bm{\gamma})^T \in \mathbbm{R}^{l+2}.    
\end{equation}

The regularity conditions for the exogenous variables make sure the model is well-defined and the estimation for the true parameter, $\bm{\beta}_0$, is consistent and asymptotically normal. We will discuss this part in Sec \ref{subsec: parameter estimation}.

%%%%%%%%%%%%%%%%%%%%%%%%%%%% 

Time series models of lag one are universally researched in the literature and are widely used in the industry, e.g. \cite{hudecova2015detection}, \cite{hudecova2016change} and \cite{kirch2015use}. Some properties of the Binomial AR(1) model can be proved by relying on the properties of Markov chains. 
%------------------------------------------------------------
\subsection{Properties of Binomial AR($1$)}
We will investigate the properties of the Binomial AR($1$) process in this subsection.

%%%%%%%%%%%%%%%%%%%%%%%%%%%% 
First, the $\psi$-irreducibility and aperiodicity of the process will be proven. By using these properties, we will then prove that the process has an invariant probability measure $\bm{\mu}$, and the proofs of strict stationarity and geometric ergodicity of the process follow. First, we introduce the definitions of the aforementioned Markov chain properties, which are from \cite{meyn1993markov}.
%%%%%%%%%%%%%%%%%%%%%%%%%%%% 

\begin{definition}
\label{def: irreducibility}
\textbf{$\psi$-irreducibility} \\
Starting from a point to a set $A$, we denote $\tau_A$ as the first time a chain reaches the set $A$. Then for a measure $\varphi$, $\varphi$-irreducibility for a Markov chain is defined as for every starting point $x \in \Omega$, 
\begin{equation}
\label{eq: def irreducibility}
    \varphi(A) >0 \Rightarrow P_x(\tau_A <\infty) >0,
\end{equation}
in which $P_x$ is the probability of an event when the starting point is $x$. If there exists a measure $\psi$ such that for every $\varphi$ satisfying (\ref{eq: def irreducibility}), $\psi(A) = 0$ implies $\varphi(A) = 0$, it's a maximal irreducibility measure and the irreducibility defined based on $\psi$ is $\psi$-irreducibility.
\end{definition}

\begin{definition}
\label{def: Aperiodicity}
\textbf{Aperiodicity} \\
For an irreducible Markov chain on a countable space, let $d$ denote the common period of the states in the state space $\Omega$, which means there exist disjoint sets $D_1,...,D_d \subseteq \Omega$ such that
\begin{equation*}
    \Omega = \bigcup_{i=1}^d D_i,
\end{equation*}
and the chain goes to $D_{i+1}$ with probability 1 if the current status belongs to $D_i$. That is, $P(X_t \in D_{i+1}|X_{t-1} = x) := P(x,D_{i+1}) = 1$, for $\forall x \in D_{i}, i = 1,...,d-1$. The chain is aperiodic if $d = 1$. The chain is strongly aperiodic if $P(x,x)>0$ for some $x \in \Omega$.
\end{definition}

% \begin{definition}
% \label{def: Positive recurrence}
% \textbf{Positive recurrence} \\
% For any state $x \in \Omega$, starting from state $x$, if the expected amount of time going back to state $x$, $\mathbbm{E}_x(\tau_x)$, is always bounded, the chain is a positive recurrent chain. For countable chains, positive Harris recurrence is equivalent to positive recurrence, so we don't differentiate the two properties here.
% \end{definition}

\begin{definition}
\label{def: Geometric ergodicity}
\textbf{Geometric ergodicity} \\
% The definition of ||.|| in a countable space is on Page 330 of [Meyn&Tweeie].
% The definition of geometric ergodicity in on Page 371.
Denote $\mu(y)$ as the unconditional probability of state $y$ and denote the n-step transition probability from $x$ to $y$ as $P^n(x,y)$. Following Theorem 13.1.2 in \cite{meyn1993markov}, for a positive Harris recurrent chain, if there exists a constant $r>1$ such that
\begin{equation*}
    \sum_{n=1}^\infty r^n ||P^n(x,\cdot)-\mu|| = \sum_{n=1}^\infty r^n (\sum_{y \in \Omega} |P^n(x,y)-\mu(y)|)<\infty
\end{equation*}
holds for any $x \in \Omega$, the process is geometric ergodic.
\end{definition}
%--------------------------

%%%%%%%%%%%%%%%%%%%%%%%%%%%% 
\begin{lemma}
\label{lemma:irreducibility&aperiodicity&recurrence}
The binomial time series $\{X_t\}$ defined as (\ref{equation: model definition1}) and (\ref{equation: model definition2}) is a $\psi$-irreducible and aperiodic Markov chain. 
%%%%%%%%%%%%%%%%%%%%%%%%%%%% 

The proof is in Section \ref{app: proof of irreducibility, aperiodicity&recurrence}. In \cite{wang2011autopersistence}, Wang and Li prove that the Binary AR(1) model is $\psi$-irreducible. In Section \ref{app: proof of irreducibility, aperiodicity&recurrence}, we follow their proof for the $\psi$-irreducibility of the Binomial AR(1). Then, given the $\psi$-irreducibility, the aperiodicity of the process is also proved. 

\end{lemma}

\begin{theorem}
\label{theorem:stationary&ergodic}
The binomial time series $\{X_t\}$ is a strictly stationary and geometrically ergodic process.
\end{theorem}
 The proof can be found in Section \ref{app: proof of stationarity and ergodicity}. 
%--------------------------------------------------------------------------------------------
\subsection{Parameter estimation}
\label{subsec: parameter estimation}
In this subsection, we first propose regularity conditions for the parameter space the true parameter vector $\bm{\beta}_0$ is in. Under the conditions, we will discuss how to estimate the true parameter vector $\bm{\beta}_0$ given $m$ observations following the model (\ref{equation: model definition1}) and (\ref{equation: model definition2}). As $m \rightarrow \infty$, the consistency and the asymptotic normality of the estimated vector, $\hat{\bm{\beta}}$, will also be proved. 

In general, suppose the distribution function of a set of random variables is known and the related regularity conditions are satisfied, then we can use maximum likelihood estimation. For time-dependent cases, especially for cases including random covariates, however, MLE can be complicated. Therefore, we take advantage of partial likelihood in this work because of its good properties in parameter estimation and its flexibility on handling covariates. Partial likelihood was first introduced in \cite{cox1975partial}. When the likelihood function contains the unknown parameters and time-dependent random covariates, the likelihood function of the observation at $t$ can be expressed as the product of the conditional likelihood of the covariates at $t$ and the conditional likelihood of the observation given the covariates. Partial likelihood only includes the conditional likelihoods of the observations and gets rid of the conditional likelihoods of the covariates.

\begin{definition}
\label{def: PL}

%%%%%%%%%%%%%%%%%%%%%%%%%%%%%%%%%%%%%%%%%%%%%%%%%%%%
Denote the observed series as $\{X_t\},t=0,1,...,m$, and denote the covariate series influencing the distribution of $\{X_t\}$ as $\{\bm{W}_t\},t=1,...,m$. 
At fixed parameter $\bm{\beta}$, as proven in Theorem \ref{theorem:stationary&ergodic}, $\{X_t\}$ is stationary. So we are able to get the unconditional distribution function for $\{X_t\}$, denoted as $f_{\bm{\beta}}(X_t)$. According to (\ref{equation: model definition1}) and (\ref{equation: model definition2}), $\{\bm{W}_t\}$ is i.i.d. with known PDF. We denote it as $h(\bm{W}_t)$. Then, according to the model definition (\ref{equation: model definition1}) and (\ref{equation: model definition2}), the distribution of $X_t$ is known once we know $X_{t-1}$ and $\bm{W}_t$. We denote the conditional distribution function as $f_{\bm{\beta}}(X_t|X_{t-1}, \bm{W}_t)$. The full likelihood function is
\begin{equation*}
    f(\bm{\beta};X_0,\bm{W}_1,X_1,...,\bm{W}_m,X_m) = f_{\bm{\beta}}(X_0)\Big[\prod_{t=1}^m h(\bm{W}_t)\Big] \Big[\prod_{t=1}^m f_{\bm{\beta}}(X_t|X_{t-1}, \bm{W}_t)\Big].
\end{equation*}
Then, the partial likelihood is defined as the second product on the right hand side.
\begin{equation*}
    \text{PL}(\bm{\beta}) = \prod_{t=1}^m f_{\bm{\beta}}(X_t|X_{t-1}, \bm{W}_t).
\end{equation*}
\end{definition}
%%%%%%%%%%%%%%%%%%%%%%%%%%%%%%%%%%%%%%%%%%%%%%%%%%%%

In \cite{wong1986theory}, general conditions are proposed for consistency and asymptotic normality of maximum partial likelihood estimator (MPLE), and a method for the calculation of the efficiency of the estimator is studied. In \cite{slud1994partial} and \cite{kedem2005regression}, some specific regularity conditions of the parameters in logistic regression and binary time series, respectively, and the asymptotic behavior of the MPLE under the conditions are proposed. Then, in \cite{fokianos2004partial}, the authors expand the specific regularity conditions to time series following generalized linear models and find the asymptotic behavior of the MPLE. 
% In \cite{hudecova2013structural}, \cite{kirch2015detection} and \cite{fokianos2014retrospective}, the estimated parameter vector of the binary time series is found as the optimal solution that maximizes the partial log-likelihood function. 
Following the definition in \cite{fokianos2004partial}, we can write out the partial likelihood of Binomial AR(1).

For Binomial AR(1), 
\begin{equation*}
    % \label{equation: PL for BinAR(1)}
    \text{PL}(\bm{\beta}) = \prod_{t=1}^m {n \choose {X_t}} \pi_t(\bm{\beta})^{X_t}(1-\pi_t(\bm{\beta}))^{n-X_t} \propto \prod_{t=1}^m \pi_t(\bm{\beta})^{X_t}(1-\pi_t(\bm{\beta}))^{n-X_t},
\end{equation*}
where $\bm{\beta} = (\varphi_0,\varphi_1, \bm{\gamma})^T \in \mathbbm{R}^{l+2}$. 

The partial score vector is defined as
\begin{equation*}
   \begin{aligned}
    \text{PSV}_m(\bm{\beta}) &= \triangledown \log\text{PL}(\bm{\beta}). \\
\end{aligned} 
\end{equation*}

\begin{equation*}
\begin{aligned}
    & \frac{\partial}{\partial \varphi_0}\log\text{PL}(\bm{\beta}) \\
    = &\frac{\partial}{\partial \varphi_0} \Big(\sum_{t=1}^m X_t \bm{\beta}^T\mathbbm{Z}_{t-1}+n\log(1-\pi_t(\bm{\beta}))\Big) \\
    =& \frac{\partial}{\partial \varphi_0} \Big(\sum_{t=1}^m X_t \bm{\beta}^T\mathbbm{Z}_{t-1}+n \log \frac{\exp(-\bm{\beta}^T\mathbbm{Z}_{t-1})}{1+\exp(-\bm{\beta}^T\mathbbm{Z}_{t-1})}\Big) \\
    =& \frac{\partial}{\partial \varphi_0} \Big(\sum_{t=1}^m (X_t-n)\bm{\beta}^T\mathbbm{Z}_{t-1}-n \log(1+\exp(-\bm{\beta}^T\mathbbm{Z}_{t-1})\Big) \\
    = & \sum_{t=1}^m \Big(X_t-n+n\frac{\exp(-\bm{\beta}^T\mathbbm{Z}_{t-1})}{1+\exp(-\bm{\beta}^T\mathbbm{Z}_{t-1})}\Big) \\
    = &\sum_{t=1}^m X_t-n \pi_t(\bm{\beta}).
\end{aligned}
\end{equation*}
Similarly
\begin{equation*}
\begin{aligned}
     & \frac{\partial}{\partial \varphi_1}\log\text{PL}(\bm{\beta}) \\
    = & \sum_{t=1}^m \Big(X_{t-1}(X_t-n)+n X_{t-1}\frac{\exp(-\bm{\beta}^T\mathbbm{Z}_{t-1})}{1+\exp(-\bm{\beta}^T\mathbbm{Z}_{t-1})}\Big) \\
    = &\sum_{t=1}^m X_{t-1}(X_t-n \pi_t(\bm{\beta})),
\end{aligned}
\end{equation*}
and with respect to the coefficient of the $j$th element of $\bm{W}_t$,
\begin{equation*}
\begin{aligned}
     & \frac{\partial}{\partial \gamma_j}\log\text{PL}(\bm{\beta}) \\
    = &\sum_{t=1}^m W_{t,j}(X_t-n \pi_t(\bm{\beta})).
\end{aligned}
\end{equation*}
The partial score vector can be written as 
\begin{equation}
\label{eq: closed-form expression of G}
    \begin{aligned}
    \text{PSV}_m(\bm{\beta}) & = \sum_{t=1}^m \mathbbm{Z}_{t-1}(X_t-n \pi_t(\bm{\beta})) \\
    & \overset{\Delta}{=} \sum_{t=1}^m G(X_t,\bm{\beta}|\mathcal{C}_{t}),
    \end{aligned}
\end{equation}
where $\mathcal{C}_{t} = \sigma(\bm{W}_t,X_{t-1},\bm{W}_{t-1},...,\bm{W}_1,X_0)$.
For the sake of simplicity, we denote $G(X_t,\bm{\beta}|\mathcal{C}_{t})$ as $G(X_t,\bm{\beta})$ onward. According to (\ref{eq: closed-form expression of G}), $G$ is a continuous function with respect to both $\{\bm{W}_t\}$ and $\{X_t\}$. Then, from Proposition 2.1.1 in \cite{straumann2005estimation}, we can conclude that with a fixed $\bm{\beta}$, $G(X_t,\bm{\beta})$ is stationary and ergodic. Similar with the method we use to prove the strict stationarity of $\{X_t\}$ in Theorem \ref{theorem:stationary&ergodic}, as the transition probability from $G(X_t,\bm{\beta})$ to $G(X_{t+1},\bm{\beta})$ is free from $t$, $G(X_t,\bm{\beta})$ is strictly stationary.

In the upcoming Theorem \ref{prop: consistency&CLT}, we prove that the MPLE $\hat{\bm{\beta}}$ obtained by solving the equation
\begin{equation}
\label{eq: estimatebeta}
    \text{PSV}_m(\bm{\beta}) = 0
\end{equation}
is almost surely unique.
Before addressing the consistency and the central limit theorem of $\hat{\bm{\beta}}$, we first propose some regularity conditions for the parameter vector $\bm{\beta}$.

%---------------------------------------------------------------------------------------
\begin{assumption}
\label{assumption: compact}
The parameter space of $\bm{\beta}$, denoted as $\bm{B}$, is convex and compact, and the true parameter $\bm{\beta}_0 = (\varphi_{0,0},\varphi_{0,1},\bm{\gamma}_0)^T$ is in the interior of $\bm{B}$. 
\end{assumption}

% Assumption \ref{assumption: compact} is proposed so that the Binomial AR(1) process $\{X_t\}$ satisfy the following Assumption, which is proposed in \cite{fokianos2004partial} to guarantee the uniqueness/existence, consistency and asymptotic normality of $\bm{\hat{\beta}}$ for time series following GLM.

% \begin{assumption}
% \label{assumption: PMLE}
% (a) The true parameter $\bm{\beta}_0$ belongs to an open set $\bm{B}^o \in \mathbbm{R}^{l+2}$. \\
% (b) The covariate vector $\mathbbm{Z}_{t-1}$ almost surely lies in a nonrandom compact subset $ \Gamma\in\mathbbm{R}^{l+2}$, such that $P[\sum_{t=1}^m \mathbbm{Z}_{t-1} \mathbbm{Z}_{t-1}^T >0]  =1.$ In addition, $\mathbbm{Z}_{t-1}^T \bm{\beta}$ lies almost surely in the domain $\bm{H}$ of the inverse link function $h = g^{-1}$ for all $\mathbbm{Z}_{t-1} \in \Gamma$ and $\bm{\beta} \in \bm{B}.$ \\
% (c) The inverse link function $h$ is twice continuously differentiable and $|\partial h(\eta)/\partial \eta| \neq 0$ for all $\eta \in \mathbbm{R}$. \\
% (d) There is a probability measure $\nu$ on $\mathbbm{R}^{l+2}$ such that $\int_{\mathbbm{R}^{l+2}} \mathbbm{Z}\mathbbm{Z}^T \nu(\text{d} \mathbbm{Z})$ is positive definite, and such that for all Borel sets $A \in \mathbbm{R}^{l+2}$, 
% \begin{equation*}
%     \frac{1}{m} \sum_{t=1}^m \mathbbm{1}_{[\mathbbm{Z}_{t-1} \in A]} \overset{P}{\rightarrow} \nu(A),
% \end{equation*}
% at the true parameter $\bm{\beta}_0.$
% \end{assumption}
%-------------------------------------------------------------------------------------
\begin{theorem}
\label{prop: consistency&CLT}
Under Assumption \ref{assumption: compact}, \\
(a) The MPLE $\hat{\bm{\beta}}$ is almost surely unique for all sufficiently large $m$.\\
(b) $\hat{\bm{\beta}} \overset{P}{\rightarrow}\bm{\beta}_0$ as $m \rightarrow \infty.$\\
(c) $\sqrt{m}(\hat{\bm{\beta}}-\bm{\beta}_0) \overset{\mathcal{D}}{\rightarrow} N(\bm{0},(-\mathbbm{E}\triangledown G(X_1,\bm{\beta}_0))^{-1})$ as $m \rightarrow \infty$.  
\end{theorem}

The proof can be found in Section \ref{app: proof of consistency and asymptotic normality}. Relying on the properties of the exogenous variables $\{\bm{W}_t\}$ (bounded with known distribution function, details can be found in the discussion following (\ref{equation: model definition1})) and Assumption \ref{assumption: compact}, we can show that the Binomial AR(1) model satisfies the regularity conditions proposed in \cite{fokianos2004partial}, which guarantee the uniqueness/existence, consistency and asymptotic normality of $\bm{\hat{\beta}}$ for time series following GLM. We prove the regularity conditions in Section \ref{app: proof of consistency and asymptotic normality}. The properties of $\hat{\bm{\beta}}$ then can be proven. 

We give the closed-form expression of $\triangledown G(X_t,\bm{\beta})$ here. From (\ref{equation: model definition1}) we know that 
\begin{equation*}
    \log(\frac{n \pi_t(\bm{\beta})}{n-n\pi_t(\bm{\beta})}) = \bm{\beta}^T \mathbbm{Z}_{t-1}. 
\end{equation*}
So \begin{equation*}
    \pi_t(\bm{\beta}) = \frac{1}{1+\exp(-\bm{\beta}^T \mathbbm{Z}_{t-1})}.
\end{equation*}
The derivative of $\pi_t(\bm{\beta})$ over the $j^{\text{th}}$ element of $\bm{\beta}$ is
\begin{equation*}
    \frac{\partial}{\partial \bm{\beta}_j} \pi_t(\bm{\beta})= -\frac{\exp(-\bm{\beta}^T\mathbbm{Z}_{t-1})}{(1+\exp(-\bm{\beta}^T\mathbbm{Z}_{t-1}))^2} \mathbbm{Z}_{t-1,j},
\end{equation*}
where $\mathbbm{Z}_{t-1,j}$ denotes the $j^{\text{th}}$ element in $\mathbbm{Z}_{t-1}.$
Then, according to the expression of $G$ shown in (\ref{eq: closed-form expression of G}), the $(i,j)^{\text{th}}$ element of $\triangledown G(X_t,\bm{\beta})$ is
\begin{equation*}
\begin{aligned}
    \frac{\partial}{\partial \bm{\beta}_j}G_i(X_t,\bm{\beta}) &= \frac{\partial}{\partial \bm{\beta}_j}(X_t-n\pi_t(\bm{\beta}))\mathbbm{Z}_{t-1,i}\\
    &= -n\mathbbm{Z}_{t-1,i}\frac{\partial}{\partial \bm{\beta}_j}\pi_t(\bm{\beta})\\
    &= -n \mathbbm{Z}_{t-1,i}\mathbbm{Z}_{t-1,j}\frac{\exp(-\bm{\beta}^T\mathbbm{Z}_{t-1})}{(1+\exp(-\bm{\beta}^T\mathbbm{Z}_{t-1}))^2}.
\end{aligned}
\end{equation*}

Therefore,
\begin{equation}
\label{eq: sample information matrix and PSV}
    \triangledown G(X_t,\bm{\beta}) = -n \mathbbm{Z}_{t-1} \mathbbm{Z}_{t-1}^T \frac{\exp(-\bm{\beta}^T\mathbbm{Z}_{t-1})}{(1+\exp(-\bm{\beta}^T\mathbbm{Z}_{t-1}))^2}.
\end{equation}
In practice, due to the ergodicity of $\{X_t\}$ and the consistency of $\hat{\bm{\beta}}$, we are able to estimate the variance-covariance matrix $(-\mathbbm{E}\triangledown G(X_1,\bm{\beta}_0))^{-1}$ by $[-\sum_{t=1}^m \triangledown G(X_t,\hat{\bm{\beta}})/m]^{-1}$ when $m$ is large enough.
%----------------------------------------------------------------------------------------------------
\section{Sequential change-point detection for the Binomial AR(1) model}
\label{sec: monitoring scheme}
In this section, we will introduce the close-end monitoring scheme for the Binomial AR(1) model. Given that there are $m$ initial observations and monitoring starts from $m+1$, close-end monitoring means that the monitoring process will be terminated if there is no change point detected within the $Nm$ new observations after the initial observations, where $N$ is a fixed and predetermined constant. This monitoring scheme is completely achievable and very common in practice. Since the parameter estimation is based on the first $m$ observations, we first state the non-contamination assumption.

\begin{assumption}
\label{assumption: no contamination assumptions}
(a) There is no change point in the exogenous variable $\bm{W}_t$. \\
(b) There is no change point in the first $m$ observations $X_t$. 
\end{assumption}
Assumption \ref{assumption: no contamination assumptions} (a) indicates that we only discuss the case when any parameters in $\bm{\beta}$ change. Assumption \ref{assumption: no contamination assumptions} (b) guarantees that given the first $m$ observations, we can estimate the parameter vector $\hat{\bm{\beta}}$. 
Under Assumption \ref{assumption: no contamination assumptions}, for the first $m$ observations, by maximizing the partial likelihood introduced in Section \ref{sec: Model definition}, we get the consistent estimator $\hat{\bm{\beta}}$ satisfying 
\begin{equation}
\label{equation: G}
\begin{aligned}
    &\text{PSV}_m(\hat{\bm{\beta}}) \\
    =&\sum_{t=1}^m \mathbbm{Z}_{t-1}(X_t-n \pi_t(\hat{\bm{\beta}})) \\
=& \sum_{t=1}^m G(X_t,\hat{\bm{\beta}}) = 0.
\end{aligned}
\end{equation}

Also, with the true parameter $\bm{\beta}_0$,
\begin{equation*}
    \mathbbm{E}G(X_1,\bm{\beta}_0) = \mathbbm{E}\big[\mathbbm{E}(G(X_1,\bm{\beta}_0)|\mathbbm{Z}_0)\big] = 0.
\end{equation*}
Given new observations $X_{m+1},...,X_{m+k}$, $1\leq k \leq Nm$, denote the sum of $G(X_t,\hat{\bm{\beta}})$ as
% In Binomial AR(1), the dimension of $G$ is low because only the previous value is included. However, when $\mathbbm{Z}_{t-1}$ includes not only the previous value but also some exogenous variables affecting $\pi_t(\hat{\bm{\beta}}_0)$, the dimension of $G$, $d$, would by high. Since the test statistics need to be updated every time a new data point arises, the monitoring process would be slow and computationally expensive if the dimension $d$ is too large. Therefore, instead of directly calculating the sum of $G$, 
$$
\bm{S}(m,k) = \sum_{t=m+1}^{m+k} G(X_t,\hat{\bm{\beta}}).
$$
% can be used as the detector statistic, which is the sum of a suitable monitoring function $H \in \mathbbm{R}^{d'}$ over $X_{m+1}$ to $X_{m+k}$, with $d'\leq d$. The monitoring function $H$ can be different from the estimating function $G$ as long as $\mathbbm{E}H(X_1,\bm{\beta}_0) = 0$ holds, and $H$ satisfies some regularity conditions that will be discussed later, and maintains the features of the series revealed by the sum of $G$, e.g. trend of the series and occurrence of extreme cases. When we set $H=G$, $\bm{S}(m,k)$ would just be the sum of the partial score vector from $m+1$ to $m+k$.  
% Due to the relatively low dimension of $H$, the efficiency of updating $\bm{S}(m,k)$ would be improved. It's pointed out in $\cite{kirch2015use}$ that, however, choosing a monitoring function with lower dimension may lead to undetectability for some alternatives. In the following paper, we use $G$ as the monitoring function. 
Then under the null hypothesis, due to the consistency of $\hat{\bm{\beta}}$, $\bm{S}(m,k)$ is expected to be reasonably close to 0; otherwise, suppose a change point appears at $k^*$ and the process never comes back to the original one, $\bm{S}(m,k), k>k^*,$ would be gradually far from 0 as more and more data arise after $X_{m+k^*}$.

Following this idea, suppose that $\{X_t\},t=1,2,...,m,$ follows model (\ref{equation: model definition1}) and (\ref{equation: model definition2}) with parameter $\bm{\beta}_0$. For the new observations, we denote $\{X_t\}, t=m+1,m+2,...,m+Nm$, follows the model with parameter $\bm{\beta}_t$, then based on the close-end monitoring scheme, we can propose test statistics with respect to $\bm{S}(m,k)$ to test the hypotheses
\\

$H_0: \bm{\beta}_t = \bm{\beta}_0 \text{ for all }t \in \{m+1,m+2,...,m+Nm\},$ 

 v.s.  
 
$H_a: \exists k^*\text{ such that } \bm{\beta}_t = \bm{\beta}_0 \text{ for } m+1\leq t< m+k^* <m+Nm \text{, and }\bm{\beta}_t \neq \bm{\beta}_0 \text{ for } m+k^*\leq t\leq m+Nm.$
\\

In this paper, we extend the monitoring scheme proposed in \cite{kirch2015use} to Binomial AR(1) sequential change-point detection that the null is rejected when 
\begin{equation}
\label{equation: test statistic}
    w^2(m,k)\bm{S}(m,k)^T\bm{A}\bm{S}(m,k) \geq c,
\end{equation}
where $w(m,k)$ is a weight function and $\bm{A}$ is a known symmetric positive definite matrix customized based on the research interest. $\bm{A}$ is used to adjust the magnitude of $\bm{S}(m,k)^T \bm{S}(m,k)$. Analogous to the Lagrange multiplier test, a common choice for $\bm{A}$ is the inverse of the information matrix of the series, e.g. \cite{huvskova2005monitoring} and \cite{hudecova2013structural}. Due to ergodicity of the process and the consistency of $\hat{\bm{\beta}}$, it can be estimated by the inverse of the average of the sample information matrix of the first $m$ observations, $[-\sum_{t=1}^m \triangledown G(X_t,\hat{\bm{\beta}})/m]^{-1}$. If $\bm{A}$ is the identity matrix, then the test is based on $\bm{S}(m,k)^T \bm{S}(m,k),$ which is the $L^2$-norm of the score vector $\bm{S}(m,k)$. As the number of new data points increases, the variance of $\bm{S}(m,k)$ increases. Therefore, the weight function, $w(m,k)$, is introduced to balance the extent of the oscillation of $\bm{S}(m,k)^T\bm{A}\bm{S}(m,k)$ in the sequential detection. Adding $w(m,k)$ to the test statistic, the asymptotic distribution of the test statistic can be obtained. The regularity conditions of the weight function will be discussed in Section \ref{sec: null}. After the asymptotic distribution of the test statistic under $H_0$ is obtained, the threshold $c$ for certain significance level can be determined as the corresponding quantile of the distribution. Ideally, under $H_a$, (\ref{equation: test statistic}) will eventually hold with probability 1 with a large enough $m$ (see Section \ref{sec: power analysis} for further discussion). 

In terms of the close-end monitoring scheme, the null hypothesis would be rejected as soon as the test statistic exceeds the threshold $c$. Mathematically, for the close-end monitoring, we want to achieve
\begin{equation*}
    \underset{m\rightarrow \infty}{\lim}P(\underset{1\leq k \leq Nm}{\sup}w^2(m,k)\bm{S}(m,k)^T\bm{A}\bm{S}(m,k) \geq c|H_0)  = \alpha,
\end{equation*}
and
\begin{equation*}
    \underset{m\rightarrow \infty}{\lim}P(\underset{1\leq k \leq Nm}{\sup}w^2(m,k)\bm{S}(m,k)^T\bm{A}\bm{S}(m,k) \geq c|H_a)  = 1.
\end{equation*}
%The stopping rule $\tau(m)$ for the close-end monitoring is defined as
%$$
%\tau(m) = \begin{cases}   \inf\{1\leq k<Nm: w^2(m,k)\bm{S}(m,k)^T\bm{A}\bm{S}(m,k) \geq c\} \\
                          %\infty, \text{ when } w^2(m,k)\bm{S}(m,k)^T\bm{A}\bm{S}(m,k) < c \text{ for all } 1 \leq k < Nm.\\
                         %\end{cases}
%$$
%$$
%\tau(m) =  \inf\{k \geq 1: w^2(m,k)\bm{S}(m,k)^T\bm{A}\bm{S}(m,k) \geq c\}.
%$$

%------------------------------------------------
\section{Regularity conditions and the asymptotic results under the null hypothesis}
\label{sec: null}
In this section, we derive asymptotic results of $w^2(m,k)\bm{S}(m,k)^T\bm{A}\bm{S}(m,k)$ under the null hypothesis and give the regularity conditions required for the results.
%----------------------------------------------------
\cite{kirch2015use} proposes some regularity conditions on the weight function $\omega(m,k)$ in the monitoring scheme. In this paper, we use the same regularity conditions as follows 
\begin{assumption}
\label{assumption: weight function}
The weight function satisfies the form
$$
\omega (m,k) = m^{-\frac{1}{2}}\Tilde{\omega}(m,k),
$$
where
$$
\Tilde{\omega}(m,k) = \begin{cases} \rho(k/m),& \text{ for }k \geq a_m \text{ with } a_m/m \rightarrow0\\
                                    0, &\text{ for } k<a_m.\end{cases}
$$
The function $\rho(.)$ is continuous and satisfies 
$$
\underset{t\rightarrow 0} {\lim} t^{\gamma}\rho(t) <\infty \text{ for some } 0\leq \gamma < \frac{1}{2},
$$
and
$$
\underset{t \rightarrow \infty}{\lim} t\rho(t)<\infty.
$$
The tuning parameter $\gamma$ in the weight function determines the sensitivity of the method. The relationship will be further discussed in Section \ref{sec: power analysis} and Section \ref{sec: real life data}.\\
\end{assumption} 

\begin{lemma}
\label{proposition: CUSUM approximation}
Under Assumption \ref{assumption: compact}-\ref{assumption: no contamination assumptions} and under $H_0$, with a weight function $\omega(m,k)$ satisfying Assumption \ref{assumption: weight function}, 
$$
\underset{k\geq 1}{\sup} \; \omega(m,k)||\sum_{t=m+1}^{m+k} G(X_t,\hat{\bm{\beta}})-(\sum_{t=m+1}^{m+k}G(X_t,\bm{\beta}_0)-\frac{k}{m}\sum_{t=1}^m G(X_t,\bm{\beta}_0))|| = o_p(1).
$$
\end{lemma}

The proof of the result can be found in Section \ref{app: proof of CUSUM approximation}. 

The proof of this lemma follows the proof for Proposition 5.2 in \cite{kirch2015use}. From the lemma we know that by taking advantage of $\omega(m,k)$, $\bm{S}(m,k) = \sum_{t=m+1}^{m+k} G(X_t,\hat{\bm{\beta}})$ can be used to estimate $\sum_{t=m+1}^{m+k}G(X_t,\bm{\beta}_0)-\frac{k}{m}\sum_{t=1}^m G(X_t,\bm{\beta}_0)$, which includes the unknown true parameter $\bm{\beta}_0$. 

\begin{proposition}
\label{proposition: wiener processes}
For any fixed $N>0$, under Assumption \ref{assumption: compact}-\ref{assumption: no contamination assumptions} and under $H_0$, as $m \rightarrow \infty$, the sequence of functions
\begin{equation*}
    \frac{1}{\sqrt{m}}\sum_{t=1}^{\floor{ms}}G(X_t,\bm{\beta}_0), \text{ for }0<s\leq N,
\end{equation*}
converges weakly to a $(l+2)$-dimensional Wiener process $W(s)$ with covariance matrix 
\begin{equation*}
    \bm{\Sigma} = \mathbbm{E}\big[G(X_1,\bm{\beta}_0)G^T(X_1,\bm{\beta}_0)\big] < \infty.
\end{equation*}
$W(s)$ is defined on $D = D[0,N]$, the space of $(l+2)$-dimensional functions that are right-continuous and have left-hand limits, with $s \in [0,N]$\footnote{The details of the definition of the space can be found \cite{oodaira1972functional}.}. 
% is a $(n+2)$-dimensional Wiener process with covariance matrix 
% \begin{equation*}
%     \bm{\Sigma} = \mathbbm{E}\big[G(X_1,\bm{\beta}_0)G^T(X_1,\bm{\beta}_0)\big] = -\mathbbm{E}\triangledown G(X_1,\bm{\beta}_0).
% \end{equation*}

%(b) There exists a Wiener process $W(k)$ with covariance matrix as (\ref{equation: wiener covariance}) so that the partial sum process $\sum_{t=1}^k (H(X_t,\bm{\beta}_0),B(\bm{\beta}_0)G(X_t,\bm{\beta}_0))$ fulfills
%$$
%\sum_{t=1}^k (H(X_t,\bm{\beta}_0),B(\bm{\beta}_0)G(X_t,\bm{\beta}_0))-W(k) = o_p(k^{\frac{1}{2}}) \text{ as } k\rightarrow \infty. 
% part c here
%$$
\end{proposition}
% Definition of n-dimensional Wiener process: page 49 of Evans 2012.
The proof is in Section \ref{app: proof of wiener process}. Here we follow the definition of $l$-dimensional Wiener processes on page 49 of \cite{evans2012introduction}, from which the standardized $l$-dimensional Wiener process, $\bm{\Sigma}^{-1/2}W(s)$, can be considered as a vector of $l$ standardized independent 1-dimensional Wiener processes. Analogous to the definition of 1-dimensional Wiener process, at $s$, the $l$-dimensional Wiener process $W(s) \sim N(\bm{0}, s \bm{\Sigma})$. Also, the increment $W(s+u)-W(s), u \geq 0$ is independent with the past values $W(t), t \leq s$. 

% In practice, we are able to simulate a $W(s)$ by generating $\lfloor ms \rfloor$ independent normal random variables following $N(\bm{0}, \bm{\Sigma})$ and calculating the cumulative sum over $m$.  Since we have proved the covariance matrix of the Wiener process, $\Sigma$, is bounded, due to the ergodicity of $G(X_t,\bm{\beta}_0)$ and the consistency of $\hat{\bm{\beta}}$, in practice, $\Sigma$ can be estimated by MCMC generating samples from the Binomial AR(1) model with parameter $\hat{\bm{\beta}}$.

\begin{theorem}
\label{theorem: null}
With the weight function $\omega(m,k)$ satisfying Assumption \ref{assumption: weight function}, under $H_0$, for any fixed symmetric positive definite matrix $\bm{A}$, 
\begin{equation*}
\begin{aligned}
    \underset{m \rightarrow \infty}{\lim}&P\left(\underset{1\leq k \leq Nm}{\sup} \omega(m,k)^2\bm{S}(m,k)^T\bm{A}\bm{S}(m,k) \leq c    \right) \\
     = &P \left( \underset{0<s\leq N}{\sup} \rho^2(s)(W_1(s)-sW_2(1))^T\bm{A}(W_1(s)-sW_2(1))\leq c \right),
\end{aligned}
\end{equation*}
where $W_1(.)$ and $W_2(.)$ are independent Wiener processes with the same covariance matrix $\bm{\Sigma}$.
\end{theorem}
The proof is in Section \ref{app: proof of the null}. It is similar with the proof for Theorem 3.1 of \cite{kirch2015detection}. 
% ---------------------------------------------

\section{Power analysis of the hypothesis test}
\label{sec: power analysis}
In this section, we will prove that when a change point occurs, as $m \rightarrow \infty$, the probability to reject the null hypothesis is 1. 
We first propose some assumptions for the change point and the process after the change point.
\begin{assumption}
\label{assumption: power analysis}
(a) The change point happens at $m+k^*$, where $k^* = \lfloor m \nu \rfloor$, $l<\nu<N$. $l$ and $N$ are fixed and free from $m$. \\
(b) There exists a $x_0$, $\nu<x_0<N$, and a compact neighborhood of $x_0$, $U_{x_0}$, such that $\frac{\lfloor mx_0 \rfloor}{m} \in U_{x_0}$ and $\underset{x \in U_{x_0}}{\inf}\rho(x)>0$, where $\rho(.)$ is the second term of the weight function $
w(m,k)$. \\
(c) Denote the process after the change point as $\{X_t^*\}$, then there exists $\bm{E}_{G,H_a}$ such that
\begin{equation*}
    \frac{1}{(x_0-\nu)m}\sum_{t=m+k^*+1}^{m+\lfloor mx_0 \rfloor}G(X^*_t,\hat{\bm{\beta}}) \overset{P}{\rightarrow}\bm{E}_{G,H_a},
\end{equation*}
and $\bm{A}^{1/2}\bm{E}_{G,H_a} \neq \bm{0}$.
\end{assumption}
(a) guarantees that the change point would not happen right after the monitoring procedure starts. (b) and (c) make sure that after the change point $k^*$, $G$ is still roughly ergodic and the test statistic can still be estimated. (c) excludes the cases that $\bm{E}_{G,H_a} = \bm{0}$ after the change point. In other words, the change should be identified by using the score vector $G$ based on the PMLE $\hat{\bm{\beta}}$ estimated based on the $m$ initial non-contaminated observations. In the meantime, it constrains the choice of $\bm{A}$ so that $\bm{A}^{1/2}\bm{E}_{G,H_a} \neq \bm{0}$. However, it can be naturally satisfied when $\bm{A}$ is a positive definite matrix, which is required in Theorem \ref{theorem: null}.
\begin{theorem}
\label{theorem: power analysis}
Under Assumption \ref{assumption: compact}-\ref{assumption: power analysis}, the power of the hypothesis test based on (\ref{equation: test statistic}) converges to 1. That is,
\begin{equation*}
    \underset{m \rightarrow \infty}{\lim} P\left(\underset{1\leq k \leq Nm}{\sup} \omega(m,k)^2\bm{S}(m,k)^T\bm{A}\bm{S}(m,k) \geq c  |H_a  \right) = 1.
\end{equation*}
We can eventually reject the null hypothesis if there exists a change point satisfying Assumption \ref{assumption: power analysis}.
\end{theorem}

The proof can be found in Section \ref{app: proof of power analysis}.
% ----------------------------------------------
\section{Simulation study}
\label{sec: simulation}
In order to verify the properties of binomial time series and the related asymptotic behaviors, we simulate a Binomial AR(1) model
\begin{equation}
    \label{eq: simulation model}
    X_t|\mathbbm{Z}_{t-1} \sim \text{Bin}(10,\pi_t(\bm{\beta}_0)), \text{ with } g(\mu_t) = g(10 \pi_t(\bm{\beta}_0)) = \mathbbm{Z}_{t-1}^T \bm{\beta}_0,
\end{equation}
where $\bm{\beta}_0 = [-1,0.1,0.4]$ and $\mathbbm{Z}_{t-1} = [1,X_{t-1}, W_t^*],$ in which $W_t \sim N(1,0.1)$ and $W_t^* = \min\{\max\{0,W_t\}, 10\}.$\footnote{We truncate the exogenous variable in order to guarantee it is bounded, which is required in the definition of the model (\ref{equation: model definition1}) and (\ref{equation: model definition2})). The truncation is customized based on the scale of the variable and the research interest. When we set up a small enough lower bound and a large enough upper bound, the truncated exogenous variable can be unchanged.} \\
\subsection{Consistency and asymptotic normality of $\hat{\bm{\beta}}$}
\label{subsec: simulation-consistency and normality}

In order to check the consistency of $\hat{\bm{\beta}}$, we implement three experiments with increasing numbers of observations. For each experiment, we generate 100 samples following (\ref{eq: simulation model}). The three experiments include samples with the number of observations $m = 500, 1000, 1500$, respectively. We are able to estimate $\hat{\bm{\beta}}$ based on every sample. Under consistency, $\hat{\bm{\beta}}$ should on average be closer to the true parameter $\bm{\beta}_0$ as the number of observations increases. Therefore, for each experiment, we summarize the mean square error (MSE) of the $i$th element of the PMLEs obtained from the 100 samples, 
\begin{equation*}
    \text{MSE}_i = \frac{1}{100} \sum_{k=1}^{100}(\hat{\bm{\beta}}_{k,i}-\bm{\beta}_{0,i})^2
    ,
\end{equation*}
in order to check the consistency of $\hat{\bm{\beta}}.$ The complete results are listed in Table \ref{table: consistency}. As $m$ increases, the MSEs for all the three parameters decrease, which validates the consistency of $\hat{\bm{\beta}}.$
\begin{table}[!h]
\centering
\begin{tabular}{l|llll}
\hline
MSE        & $\beta_{0,0}$  & $\beta_{0,1}$ & $\beta_{0,2}$   \\ \hline
$m = 500$ & 0.11564 & 0.00037     & 0.10883     \\
$m = 1000$ & 0.08002  & 0.00029    & 0.07598    \\
$m = 1500$ & 0.06307   & 0.00023    & 0.05937     \\ \hline
\end{tabular}
\caption{MSEs of the PMLEs from the three experiments.}
\label{table: consistency}
\end{table}

\begin{figure}[!h]
    \centering
    \includegraphics[width=0.4\textwidth]{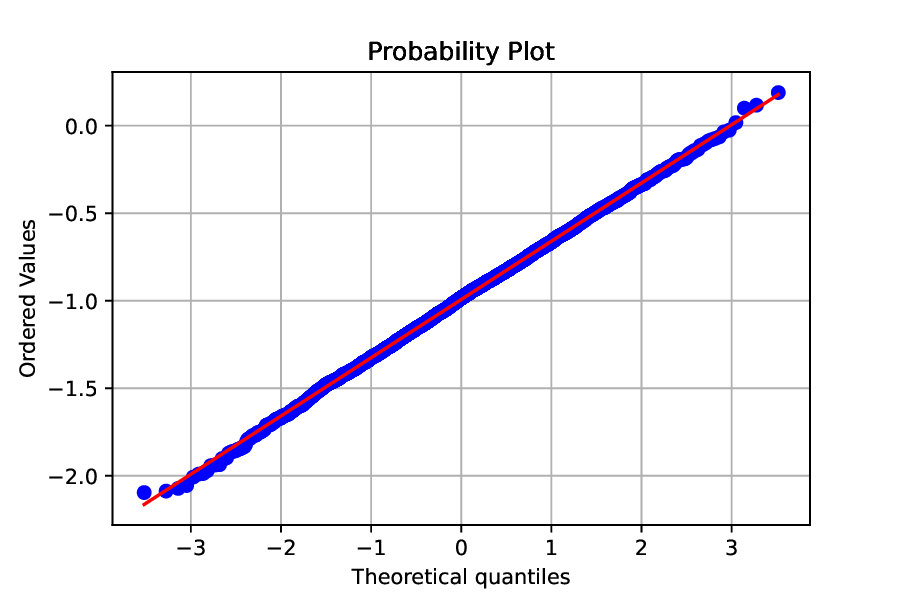}
    \includegraphics[width=0.4\textwidth]{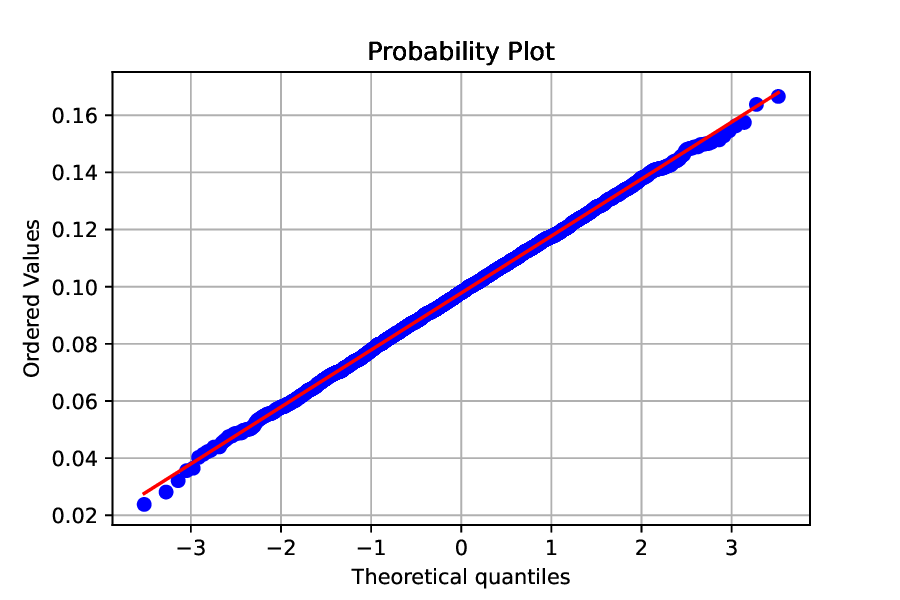} 
    \includegraphics[width=0.4\textwidth]{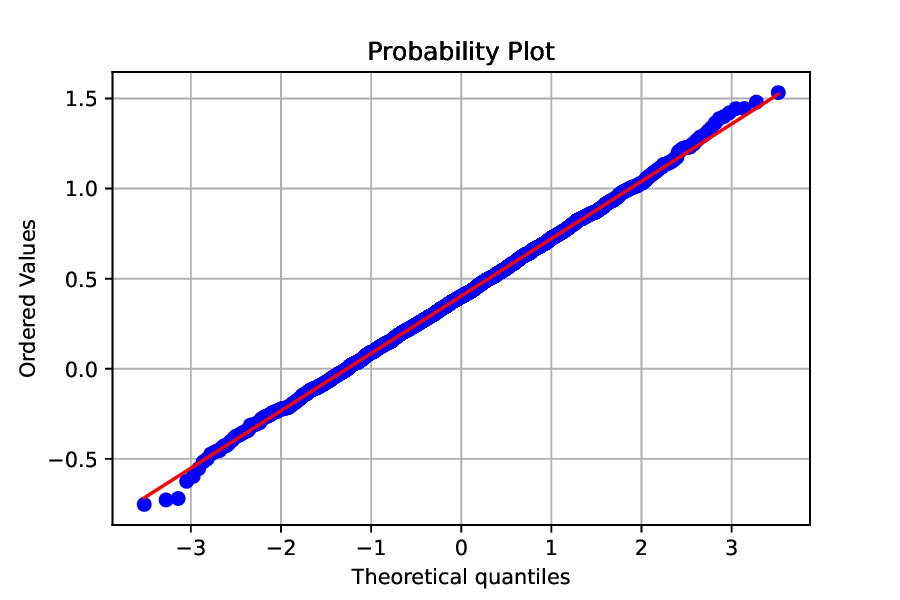}
    \caption{The Q-Q plots of the four estimated parameters based on the 1000 replications. The true parameter vector is [-1, 0.1, 0.4].}
\label{fig: qqplots}
\end{figure}

We generate 3,000 samples to assess the asymptotic normality of the estimated parameter vector. For each sample, there are 401 observations $(X_0,X_1,...,X_{400})$ generated from the model above. That is, $m=400$. Then, for each sample, we can get the PMLE $\bm{\hat{\beta}}$. The average of the 3,000 estimated $\hat{\bm{\beta}}$ is $[-0.9931,0.0980, 0.4036]$, which is close to the true value. Moreover, Henze-Zirkler test \cite{henze1990class}, which is used to assess multivariate normality, is conducted on the 3,000 estimated $\bm{\hat{\beta}}$ and the $p$
-value is 0.50. Therefore, based on the test, we can conclude that $\bm{\hat{\beta}}$ follows a normal distribution. For each parameter, we also get the quantile-quantile plot (Q-Q plot) for visualizing the marginal distribution of the parameters. Theoretically, the marginal distribution of each parameter is a normal distribution. From the four Q-Q plots shown in Figure \ref{fig: qqplots}, we can see that the marginal distributions can be considered as normal distributions.

\subsection{Asymptotic results under the null}
\label{sec: Asymptotic results under the null}
We use the same model and the same data generating process in Section \ref{subsec: simulation-consistency and normality} to check the validity of the asymptotic distribution under the null hypothesis proposed in Theorem \ref{theorem: null}. We first estimate the theoretical information matrix by generating a sequence of 10,000 observations $X_t, t=1,2,...,10000,$ and calculate $\bm{\hat{\Sigma}}_0 = \sum_{t=1}^{10000} G(X_t,\hat{\bm{\beta}})G(X_t,\hat{\bm{\beta}})^T/10000$, of which $\hat{\bm{\beta}}$ is the PMLE estimated based on the 10,000 observations. $\bm{\hat{\Sigma}}_0$ can be considered close to the theoretical information matrix due to the ergodicity of the process and the consistency of $\hat{\bm{\beta}}$. We also assign the inverse of the theoretical information matrix to the $\bm{A}$ in the test statistic, i.e. $\bm{A} = \bm{\hat{\Sigma}}^{-1}_0$. We choose the weight function
\begin{equation}
    \label{eq: definition of the weight function}
    \omega(m,k, \gamma) = m^{-\frac{1}{2}}(1+\frac{k}{m})^{-1}(\frac{k}{m+k})^{-\gamma}, 0\leq \gamma< \frac{1}{2},
\end{equation}
for the simulation. $\omega(m,k, \gamma)$ is a standard weight function in the literature, such as \cite{horvath2004monitoring} and \cite{horvath2007sequential}, and it can be easily verified that $\omega(m,k, \gamma)$ satisfies Assumption \ref{assumption: weight function}. Following the notations in Assumption \ref{assumption: weight function}, we denote 
\begin{equation*}
    \rho(s,\gamma) = s^{-\gamma}(s+1)^{\gamma-1},
\end{equation*}
then 
\begin{equation*}
    \omega(m,k, \gamma) = m^{-\frac{1}{2}}\rho(k/m, \gamma), 0\leq \gamma< \frac{1}{2}.
\end{equation*}
With a fixed $\gamma$ and the significance level $\alpha$, the threshold $c(\gamma,\alpha)$ can be determined so that 
$$P(\underset{0< s \leq N}{\sup}\rho^2(s,\gamma)(W_1(s)-sW_2(1))^TA(W_1(s)-sW_2(1))\leq c(\gamma,\alpha)) = 1-\alpha.$$

In order to investigate the influence of $\gamma$ on the sensitivity of the change point detection approach, we choose $\gamma = 0, 0.25, 0.4$ for the simulation. We first compute the threshold $c(\gamma,\alpha)$ for different $(\gamma,\alpha)$ when $N = 3$. For $N = 3$ and $\omega(m,k,\gamma)$ with each $\gamma$, 10,000 functions $\rho^2(s,\gamma)(W_1(s)-sW_2(1))^T\bm{A}(W_1(s)-sW_2(1)), 0<s\leq N$ are generated, and $\hat{\bm{\Sigma}}_0$ is used as the estimated covariance matrix of $W_1(.)$ and $W_2(.)$. Following the generating method we mentioned in Proposition \ref{proposition: wiener processes}, we choose $m = 1000$ and generate $N*m = 3000$ independent normal random vectors following $N(\bm{0}, \hat{\bm{\Sigma}}_0)$. The cumulative sum of the first $k$ random variables divided by $\sqrt{m}$ simulates $W_1(k/m), k = 1,2,...,Nm.$  Similarly, we are able to simulate a $W_2(1)$ by generating a random vector following $N(\bm{0}, \hat{\bm{\Sigma}}_0)$. For every iteration, we repeat the procedure to simulate $\rho^2(s,\gamma)(W_1(s)-sW_2(1))^T\bm{A}(W_1(s)-sW_2(1))$ with $s = \frac{1}{m}, \frac{2}{m},...,\frac{Nm}{m}$, and record the supremum over $s$. Then, the $(1-\alpha)$ quantile of the supremum of the 10,000 samples is considered as the threshold $c(\gamma, \alpha)$.
In Table \ref{table:null c }, we list the threshold $c(\gamma,\alpha)$ for different $(\gamma,\alpha)$ with $N = 3$.
\begin{table}[!htb]
\centering
\begin{tabular}{lllll}
\hline
$\alpha$        & 0.1    & 0.05   & 0.025  & 0.01    \\ \hline
$\gamma = 0$    & 5.6145 & 7.2195 & 8.6995 & 10.0376  \\
$\gamma = 0.25$ & 6.9467 & 8.4285 & 9.6381 & 12.3182  \\
$\gamma = 0.4$  & 8.8090 & 10.3182 & 11.7566 & 13.7854 \\ \hline
\end{tabular}
\caption{$c(\gamma,\alpha)$ under different $\gamma$ and significant level $\alpha$}\label{table:null c }
\end{table}

\noindent
With $m=100,200,300$ and $\gamma = 0,0.25,0.4$, the empirical probability that 
$$\underset{1\leq k \leq Nm}{\sup}\omega^{2}(m,k, \gamma)\bm{S}(m,k)^T\bm{A}\bm{S}(m,k) \geq c(\gamma,\alpha)$$
is obtained based on 10,000 repetitions, and the comparisons of the empirical probabilities and the theoretical $\alpha$'s are listed in Table \ref{table:null alpha compare}.
\begin{table}[!htb]
\centering
\begin{tabular}{llllll}
\hline
\multicolumn{2}{c}{$\alpha$}                                & 0.1    & 0.05                       & 0.025  & 0.01   \\ \hline
\multicolumn{1}{c}{\multirow{3}{*}{$\gamma = 0$}} & $m=100$ & 0.1254 & 0.0618                     & 0.0314 & 0.0183 \\
\multicolumn{1}{c}{}                              & $m=200$ & 0.1162 & \multicolumn{1}{c}{0.0512} & 0.0243 & 0.0133 \\
\multicolumn{1}{c}{}                              & $m=300$ & 0.1116 & 0.0468                     & 0.0213 & 0.0099 \\ \hline
\multirow{3}{*}{$\gamma = 0.25$}                  & $m=100$ & 0.1486 & 0.0938                     & 0.0642 & 0.0339 \\
                                                  & $m=200$ & 0.1275 & 0.0730                     & 0.0435 & 0.0206 \\
                                                  & $m=300$ & 0.1250 & 0.0707                     & 0.0420 & 0.0181 \\ \hline
\multirow{3}{*}{$\gamma = 0.4$}                   & $m=100$ & 0.1868 & 0.1260                     & 0.0832 & 0.0508 \\
                                                  & $m=200$ & 0.1690 & 0.1062                     & 0.0683 & 0.0371 \\
                                                  & $m=300$ & 0.1566 & 0.0898                     & 0.0572 & 0.0291 \\ \hline
\end{tabular}
\caption{The empirical probabilities with different $m$ and $\gamma$}\label{table:null alpha compare}
\end{table}

From (\ref{eq: definition of the weight function}), we can see that $\omega(m,k, \gamma)$ is an increasing function in terms of $\gamma$. It is revealed in Table \ref{table:null c }. As $\gamma$ increases, with the same $\alpha$, the threshold $c(\gamma, \alpha)$ increases. When the weight function is multiplied to $\bm{S}(m,k)^T\bm{A}\bm{S}(m,k)$, a large $\gamma$ enlarges the fluctuation of $\bm{S}(m,k)^T\bm{A}\bm{S}(m,k)$. Therefore, in Table \ref{table:null alpha compare}, we can see that when $\gamma = 0$ and $m$ is large enough, the empirical rejection probabilities are close to the nominal ones. However, when $\gamma = 0.4$, the empirical rejection probabilities are all higher than the corresponding nominal rejection probabilities. The experiment requires more observations $m$ in order to let the empirical probabilities further approach the theoretical ones. In other words, a detector with high $\gamma$ is more sensitive. It is the trade-off of the detection method. A high $\gamma$ enables the method to have high sensitivity. When there is a change point, the method can quickly detect it (we will further discuss it in Section \ref{sec: Asymptotic results under the alternative}). However, the drawback is a high false positive alarm. 

\subsection{Asymptotic results under the alternative}
\label{sec: Asymptotic results under the alternative}
In order to examine the power of the test, we also generate processes including change points to see whether the test can detect the change point and how sensitive the test is. For $m=100,200,300$ and $N=3$, we generate 5,000 processes. For every process, the first $m+10$ data points follow the Binomial AR(1) model with total $n = 10$, $\{W_t^*\}$ follow the same distribution introduced in (\ref{eq: simulation model}) and parameter vector $[-1,0.1,0.4]$. The remaining $Nm-10$ data points follow the Binomial AR(1) model with total $n =10$ and parameter vector $[-1,0.2,0.4]$. The parameter vector is estimated based on the first $m$ observations, and then the monitoring procedure launches from $m+1$. That is, for each $m$, we generate 5,000 processes with change point at the $\text{11}^{\text{th}}$ data point after the monitoring procedure starts.  With $\gamma = 0, 0.25, 0.4$, we record the first time point $k, 1\leq k \leq Nm$ that $\omega^{2}(m,k,\gamma)\bm{S}(m,k)^T\bm{A}\bm{S}(m,k) \geq c(\gamma,0.05)$, which is the time point we reject the no change point null hypothesis with significant level 0.05, if the $k$ exists. Meanwhile, for each $m$, we record the number of processes for which the null hypothesis is rejected, which can be used to evaluate the power of the test. In this simulation study, for each $m$, all change points in the 5,000 processes are successfully detected. It provides a strong evidence supporting Theorem \ref{theorem: power analysis}. Table \ref{table: asymptotic results under the alternative} shows the average time points of the 5,000 processes when the null hypothesis is rejected by the monitoring procedure for each $m,\gamma$ combination. Since the change point occurs at the $\text{11}^{\text{th}}$ data point, we can observe from Table \ref{table: asymptotic results under the alternative} that with the same $m$, the sensitivity of the monitoring procedure increases as $\gamma$ increases. Meanwhile, as $m$ increases, from (\ref{eq: definition of the weight function}), we can see that the weight function decreases when $s$ and $\gamma$ are fixed, which leads to the decreasing of the test statistic. That explains the slowdown of the detection speed when $m$ increases.

\begin{table}[!htb]
\centering
\begin{tabular}{l|lll}
\hline
$m$            & 100 & 200 & 300 \\ \hline
$\gamma = 0$    & 30.56	& 36.45	& 41.52 \\
$\gamma = 0.25$ & 25.30 & 27.32 & 28.94  \\
$\gamma = 0.4$  & 22.31 & 22.64 & 23.36  \\ \hline
\end{tabular}
\caption{The mean of detected change point locations of the 5,000 processes for each $m$.}
\label{table: asymptotic results under the alternative}
\end{table}

From Section \ref{sec: Asymptotic results under the null} and Section \ref{sec: Asymptotic results under the alternative} we can see that the sensitivity of the detection can be adjusted by using different tuning parameter $\gamma$. If we want to have a method with high sensitivity and a false positive alarm (detecting a change point when there is not) is not expensive, we recommend to choose a high $\gamma$ (like 0.4 in the simulations). Conversely, if given prior knowledge we consider the probability of the occurrence of a change point is low, a small $\gamma$ is preferred (like 0 in the simulations).
%-------------------------------------------------------
\section{Application to Weekly Pneumonia \& Influenza Mortality Rate} 
\label{sec: real life data}
\subsection{Data}
Here, we apply the monitoring procedure to a data set recording the weekly number of states with Pneumonia \& Influenza (P\&I) mortality rates (\% of deaths due to Pneumonia \& Influenza) higher than the corresponding averages (details see below) among the six states: California, Colorado, Illinois, Florida, New York and Washington. The data set covers the time period from the 1st week of 2017 to the 5th week of 2021.

We first collect the statewide weekly P\&I mortality rate of all 50 states from the 40th week of 2013 (CDC treats the 40th week of each year as the start of the corresponding flu season) to the 5th week of 2021\footnote{https://gis.cdc.gov/grasp/fluview/mortality.html}. The fluctuation of P\&I rates reflects the extent of the spread of pneumonia and influenza. Therefore, the original data set shows a cyclical pattern with period 52 (the number of weeks per year). In order to deseasonlize the data set, for each state, we first calculate the weekly average P\&I rate from 2013 to 2016, the years without severe pandemics. We consider the average P\&I rate of each state as the standard when evaluating P\&I rates in the future. Then, by comparing the current P\&I rate with the corresponding average (same state and same week), we generate binary time series for each state from the 1st week of 2017 to the 5th week of 2021, in which 1 indicates the current rate higher than the average and 0 otherwise. We choose the binary time series of the six states: California, Colorado, Illinois, Florida, New York and Washington, and the sum of the six binary time series is a Binomial time series with total 6. Geographically, the six states can represent the nationwide flu activity trend. Meanwhile, the six states are far enough to be considered mutually independent at each time, which satisfies the assumption of a Binomial time series.

Given the aforementioned Binomial time series, we use the developed sequential change point detection method to detect abnormal P\&I rate in order to estimate the start of Covid-19 outbreak and evaluate the influence of the pandemic.

\subsection{Detection Results}
As a preliminary analysis of the data set, we calculate the AIC of the fitted Binomial AR(1) model with $\mathbbm{Z}_{t-1} = (1, X_{t-1})$, and the AIC of a simple Binomial model with a constant $\pi$ that all the observations are considered as i.i.d random variables following $\text{Bin}(6,\pi)$. By comparing the two AICs, we can evaluate the utility of the more complicated model. Also, a likelihood ratio test is conducted between the two models. The AIC of the simple Binomial model is 2228.31 and the AIC of the Binomial AR(1) model is 1209.48. The p-value of the likelihood ratio test is $7.12\times 10^{-108}$. Compared to a simple Binomial model, the Binomial AR(1) model appears to detect dependence between observations and estimate the time-varying $\pi_t$.

\begin{figure}[!h]
    \centering
    \includegraphics[width = 8cm]{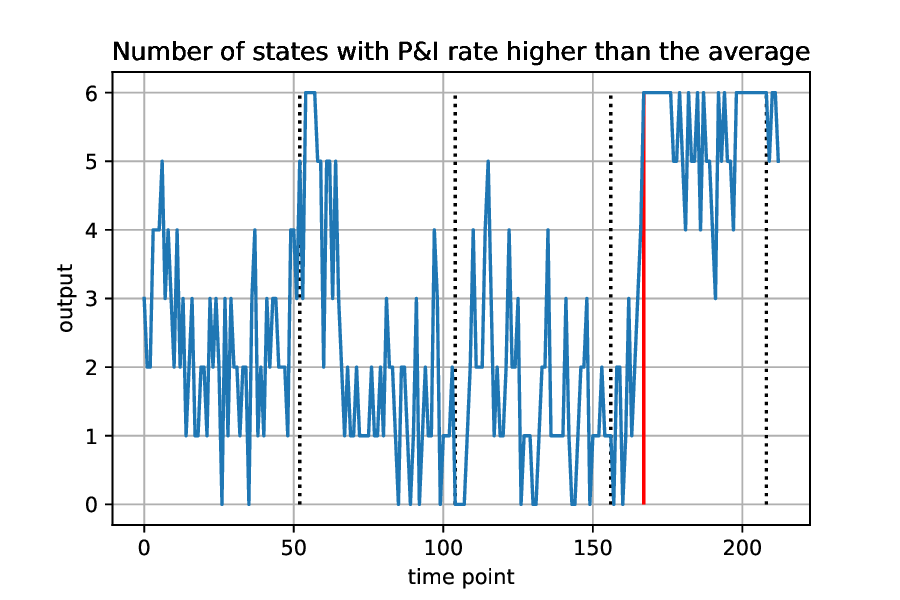}
    \caption{The number of states out of the six states exceeding the average P\&I rate from the 1st week of 2017 to the 5th week of 2021 (the dotted lines are the starts of every year, and the red solid line shows the location of 03/21/2020, the time point the p\&I rate starts to increase).}
    \label{fig:flu_binomial}
\end{figure}

From Figure \ref{fig:flu_binomial} we can see that the P\&I rate starts to increase from March 21th 2020 (the 12th week of 2020), the $167^{\text{th}}$ data point. Table \ref{Table: IUR change point} shows the change points detected by models built on different past observations. We can find that the detected change points are different due to different launch dates and different tuning parameters, which are used to adjust the sensitivity of the method. But except the only failure (when $m=100$ and $\gamma = 0.4$), all other change points are detected right after the 12th week of 2020. In the most successful case, when $m=150$ and $\gamma = 0.25, 0.4$, the change point is detected 2 points after the 12th week of 2020. The results also validate the conclusions we have from Section \ref{sec: Asymptotic results under the null} and Section \ref{sec: Asymptotic results under the alternative}. Since there indeed exists a change point, the detector with $\gamma = 0.4$ is more sensitive than the other two detectors with $\gamma = 0, 0.25$. However, in practice, it is hard to have universal criteria on the choice of $m$, as the fitted model itself changes when $m$ changes. It is helpful to choose different $m$ and compare the results.

\begin{table}[!h]
\centering
\begin{tabular}{l|lll}
\hline
Number of past observations & $m = 100$ & $m = 130$ & $m = 150$ \\
Detection launch date       & 12/8/18   & 7/6/19 & 11/23/19  \\ \hline
$\gamma = 0$                & 4/18/20  & 4/25/20  & 4/11/20   \\
$\gamma = 0.25$             & 4/11/20   & 4/18/20  & 4/4/20  \\
$\gamma = 0.4$              & 1/19/19  & 4/18/20  & 4/4/20   \\ \hline
\end{tabular}
\caption{The detected change points relying on models based on different past observations}\label{Table: IUR change point}
\end{table}

\newpage
\section{Appendix}
\subsection{Proof of Lemma \ref{lemma:irreducibility&aperiodicity&recurrence}}
\label{app: proof of irreducibility, aperiodicity&recurrence}
To prove $\psi$-irreducibility, we first prove the process is $\varphi$-irreducible. Following the notations in Definition \ref{def: irreducibility}, for the lag one binomial time series process, $\Omega = \{0,1,...,n\}$. We define the measure $\varphi$ as $\varphi(A) = \# A/(n+1)$ for any set $A \in \mathcal{B}(\Omega)$, where $\#A$ measures the number of elements in $A$. Therefore, whenever $\varphi(A)>0$, $\sum_{i=0}^n\mathbbm{1}(i \in A) >0$ holds. From the link function introduced in (\ref{equation: model definition1}), we can see that we are able to get a corresponding minimum and a maximum of $\pi_t(\bm{\beta}_0)$ once we know $X_{t-1} =  x$, and the minimum and the maximum are only determined by $x$ (free from $t$). We can't fully determine the value of $\pi_t(\bm{\beta}_0)$ without considering the output of $\bm{W}_t$. However, given that $\bm{W}_t \in [a,b]^l$, we can calculate the upper bound and lower bound of $\bm{\beta}_0^T \mathbbm{Z}_{t-1}$, which provide the corresponding upper bound and lower bound of $\pi_t(\bm{\beta}_0)$. Given Assumption \ref{assumption: compact} and the fact that every element in $\mathbbm{Z}_{t-1} = (1, X_t, \bm{W}_t)$ is bounded, $\bm{\beta}_0^T \mathbbm{Z}_{t-1}$ can't go to $-\infty$ or $\infty$. Combining it with the definition of the link function $g$, (\ref{equation: model definition1}), we can see that $\pi_t(\bm{\beta}_0)$ is unable to be exact 0 or 1. Therefore, with any fixed $i \in \Omega$, we can always find the corresponding minimum of the one-step transition probability from state $x$ to state $i$ when $x$ is known. We denote it as $\underset{-}{P_{xi}}, 0<\underset{-}{P_{xi}}<1.$

Then
$$
P_x(\tau_A < \infty) \geq P_x(\tau_A = 1) =P(X_t \in A|X_{t-1} = x) \geq \sum_{i=0}^n \underset{-}{P_{xi}} \mathbbm{1}(i \in A) >0,
$$
which implies that the process is $\varphi$-irreducible. By Theorem 4.0.1 and Prop 4.2.2 in \cite{meyn1993markov}, there exists an essentially unique maximal irreducibility measure $\psi$ on $\mathcal{B}(\Omega)$ that the process is $\psi$-irreducible. Maximal irreducibility measure means that whenever $\psi(A) = 0$, $\varphi(A) = 0$. 

Then we prove the process is aperiodic. That is, following Definition \ref{def: Aperiodicity}, we will show that $d=1$. Assume there are $d$ disjoint sets $D_1,D_2,...,D_d \subseteq \Omega$, without loss of generality, we may assume that $\varphi(D_d) > 0$. Then for any $x \in D_{d-1}, P(x,D_d) = 1$, which implies that $P(x,\cup_{i=1}^{d-1} D_i) = 0$. Due to the facts that all one-step transition probabilities of the process are positive and that the definition of $\varphi$ depends on the number of elements in the set, it follows that $\varphi(\cup_{i=1}^{d-1} D_i) = 0$, which means that $\varphi(D_d) = 1$. Therefore, $d=1$ and the process is aperiodic. 
% -----------------------------------------------------------------------------
\subsection{Proof of Theorem \ref{theorem:stationary&ergodic}}
\label{app: proof of stationarity and ergodicity}
Before proving the stationarity and geometric ergodicity of the chain, we need to first introduce $\nu_1$-small set and drift operator.

Denote the lower bound of all possible one-step transition probabilities as $\delta = \min\limits_{x,i \in \Omega}\underset{-}{P_{xi}}$, which is strictly positive as we demonstated in Section \ref{app: proof of irreducibility, aperiodicity&recurrence}. Define the measure $\nu_1$ as
$$
\nu_1(A) = \delta \times (\#A) , \text{ for } A \in \mathcal{B}(\Omega).
$$
Then for any set $C \in \mathcal{B}(\Omega)^+$, which means $C \in \mathcal{B}(\Omega)$ with $\psi(C)>0$,
$$
P(x,A) \geq \nu_1(A), \text{ for any }x \in C, A \in \mathcal{B}(\Omega).
$$
Therefore, we say that every non-empty set $C \in \mathcal{B}(\Omega)^+$ is a $\nu_1$-small set. So the state space $\Omega$ is a $\nu_1$-small set.

Following the definition on page 558 of \cite{meyn1993markov}, the drift operator is defined as
$$
\Delta V(x) = \int P(x,\text{d}y)V(y)-V(x),
$$
which is used to describe the expected drift of the function $V(\cdot)$ from state $x$ to next state. 

Then, according to Theorem 15.0.1 of \cite{meyn1993markov}, if a Markov chain is $\psi$-irreducible, aperiodic and satisfies the drift condition, 
\begin{equation}
\label{eq: drift condition}
   \Delta V(x)\leq -\beta V(x)+b\mathbbm{1}_{C}(x), x \in \Omega,
\end{equation}
where $C$ is a small set, constant $b < \infty$, $\beta >0$ and a function $V \geq 1$ finite at some one $x_0 \in \Omega$, then there exists a unique invariant measure $\bm{\mu}$.

So we first show that (\ref{eq: drift condition}) holds for the Binomial AR(1) model with $\Omega = \{0,1,...,n\}$. We have proven that the state space $\Omega$ is a small set. Let $V(x) = x+1$, $b = n+1,\beta = 1$ and $C = \Omega$, then
\begin{equation*}
    \begin{aligned}
    \Delta V(x) = & \int P(x,\text{d}y)V(y)-V(x) \\
             \leq & (\underset{y \in \Omega} {\max}V(y))-V(x) \\
              = & (n+1)-V(x) \\
              = & -V(x)+b\mathbbm{1}_{\Omega}(x), \text{ for any }x \in \Omega.
    \end{aligned}
\end{equation*}

Combining Lemma \ref{lemma:irreducibility&aperiodicity&recurrence} and the proven drift condition, we conclude that there exists a unique invariant measure $\bm{\mu} = (\mu_0,\mu_1,...,\mu_n)$ for the Binomial AR(1) model such that
\begin{equation*}
    \bm{\mu} = \bm{\mu}\bm{P}
\end{equation*}
holds, where
\begin{equation*}
\bm{P} = 
    \begin{bmatrix}
    P_{00} & P_{01}&\cdots&P_{0n} \\
    P_{10} & P_{11}& \cdots&P_{1n} \\
    \vdots & \vdots &\ddots&\vdots \\
    P_{n0}&P_{n1}&\cdots&P_{nn}
    \end{bmatrix}
\end{equation*}
is the one-step transition matrix of the time series. Given the condition that the initial distribution of $\{X_t\}$ is distributed according to $\bm{\mu}$, the binomial time series $\{X_t\}$ admits the unique invariant measure $\bm{\mu}$ such that 
\begin{equation*}
    P(X_t = k) = \mu_k, \text{ for } k = 0,1,...,n \text{ and }t \in \mathbbm{Z}.
\end{equation*}

By writing out the $n$-step transition probability by using the one-step transition probability,
\begin{equation}
\label{equation: n step transition}
    \begin{aligned}
        P^n_{ji} =& P(X_t=i|X_{t-n}=j) \\
        =& \sum_{k_1=0}^n \sum_{k_2=0}^n\cdots \sum_{k_{n-1}=0}^n P_{k_1,i}P_{k_2,.k_1}\cdots P_{j,k_{n-1}}
    \end{aligned}
\end{equation}
is a function free of $t$. That is, for any $n \in \mathbbm{Z}^+$, $P^n_{ji}$ only depends on $n,i$ and $j$.
So for any random vectors $(X_{t_1},...,X_{t_n})$ and $(X_{t_1+\tau},...,X_{t_n+\tau})$, the joint pmf of the random vectors 
\begin{equation*}
\begin{aligned}
&P(X_{t_n}= x_{t_n},X_{t_{n-1}}= x_{t_{n-1}},...,X_{t_1}= x_{t_1}) \\
=&P(X_{t_n}= x_{t_n}|X_{t_{n-1}} = x_{t_{n-1}})...P(X_{t_2}=x_{t_2}|X_{t_1}=x_{t_1})P(X_{t_1}=x_{t_1}) \\
=&P(X_{t_n+\tau}= x_{t_n}|X_{t_{n-1}+\tau} = x_{t_{n-1}})...P(X_{t_2+\tau}=x_{t_2}|X_{t_1+\tau}=x_{t_1})P(X_{t_1+\tau}=x_{t_1}) \\
 = & P(X_{t_n+\tau}= x_{t_n},X_{t_{n-1}+\tau}= x_{t_{n-1}},...,X_{t_1+\tau}= x_{t_1}), \text{ for any } \tau \in \mathbbm{Z},
\end{aligned}
\end{equation*}
by the facts that $X_{t_1}$ and $X_{t_1+\tau}$ are distributed following the same unconditional probability (row) vector $\bm{\mu}$ and the conditional distributions for each pair of $X_{t_{i+1}}|X_{t_i}$ and $X_{t_{i+1}+\tau}|X_{t_{i}+\tau}$, as defined in (\ref{equation: n step transition}), are the same, for $i=1,2,...,n-1$. Therefore, the process is strictly stationary. 

Moreover, as shown as a $\psi$-irreducible and aperiodic Markov chain, the geometric ergodicity of the chain follows after the drift condition (\ref{eq: drift condition}) is proven.

\subsection{Proof of Theorem \ref{prop: consistency&CLT}}
\label{app: proof of consistency and asymptotic normality}

In \cite{fokianos2004partial}, the authors propose the following regularity conditions.\\
A(a). The true parameter $\bm{\beta}_0$ belongs to an open set $\bm{B}^o \subset \mathbbm{R}^{l+2}$. \\
A(b). The covariate vector $\mathbbm{Z}_{t-1}$ almost surely lies in a nonrandom compact subset $ \Gamma\subset\mathbbm{R}^{l+2}$, such that $P[\sum_{t=1}^m \mathbbm{Z}_{t-1} \mathbbm{Z}_{t-1}^T >\bm{0}]  =1.$ In addition, $\bm{\beta}^T\mathbbm{Z}_{t-1}$ lies almost surely in the domain $\bm{H}$ of the inverse link function $h = g^{-1}$ for all $\mathbbm{Z}_{t-1} \in \Gamma$ and $\bm{\beta} \in \bm{B}.$ \\
A(c). The inverse link function $h$ is twice continuously differentiable and $|\partial h(\eta)/\partial \eta| \neq 0$ for all $\eta \in \mathbbm{R}$. \\
A(d). There is a probability measure $\nu$ on $\mathbbm{R}^{l+2}$ such that $\int_{\mathbbm{R}^{l+2}} \mathbbm{Z}\mathbbm{Z}^T \nu(\text{d} \mathbbm{Z})$ is positive definite, and such that for all Borel sets $A \subset \mathbbm{R}^{l+2}$, 
\begin{equation*}
    \frac{1}{m} \sum_{t=1}^m \mathbbm{1}_{[\mathbbm{Z}_{t-1} \in A]} \overset{P}{\rightarrow} \nu(A),
\end{equation*}
at the true parameter $\bm{\beta}_0.$

Under the conditions, the uniqueness/existence and the consistency of the PMLE $\hat{\bm{\beta}}$ hold and the asymptotic normality 
\begin{equation*}
    \sqrt{m}(\hat{\bm{\beta}}-\bm{\beta}_0) \overset{\mathcal{D}}{\rightarrow} N(\bm{0},(-\mathbbm{E}\triangledown G(X_1,\bm{\beta}_0))^{-1})
\end{equation*}
holds, as $m \rightarrow \infty$. So we first check that the model following (\ref{equation: model definition1}) and (\ref{equation: model definition2}) satisfies the conditions. 

A(a) can be satisfied based on Assumption \ref{assumption: compact}. 

As shown in (\ref{equation: model definition1}), $g(x) = \log (x/(n-x))$. The inverse link function defined in A(b), $h(\eta) = n/(1+e^{-\eta}).$ So the first derivative $\partial h/ \partial \eta = ne^{-\eta}/(1+e^{-\eta})^2$ and $|\partial h/ \partial \eta|\neq 0$ for all $\eta \in \mathbbm{R}$. Moreover, 
\begin{equation*}
\partial^2 h/\partial \eta^2 = \frac{n e^{-2\eta}-ne^{-\eta}}{(1+e^{-\eta})^3}
\end{equation*}
is continuous. Together with Assumption \ref{assumption: compact}, A(c) and the second part of A(b) can be proven. For the first part of A(b), we can always transform the regressor $\mathbbm{Z}_{t-1}$ to $\mathbbm{Z}_{t-1}^* = (1, X_{t-1}, \bm{W}^*_{t})$ where $\bm{W}^*_{t} = \max\{ \bm{W}_{t},\bm{W}_{t}-(\bm{a}+\bm{\epsilon})\},$ where $\bm{a}+\bm{\epsilon}$ is the $l$-dimensional vector with every element equal to $a+\epsilon$. $a$ is the known lower bound of $\bm{W}_t$, and $\epsilon>0$ can be a customized and known constant. By doing the transformation, all elements in $\mathbbm{Z}_{t-1}^*$ are non-negative and the first part of A(b) holds. Meanwhile, since $\bm{a}+\bm{\epsilon}$ is known, the parameter vector for the transformed regressor is identifiable and still possess the properties of PMLE. Therefore, under transformation, Binomial AR(1) model satisfies A(b). Then, based on Theorem \ref{theorem:stationary&ergodic} and the property of $\bm{W}_t$ that they are i.i.d. random variables, A(d) can be naturally satisfied. Therefore, following Theorem 1 in \cite{fokianos2004partial}, \\
(a) the MPLE $\hat{\bm{\beta}}$ is almost surely unique for all
sufficiently large $m$, \\
(b) $\hat{\bm{\beta}} \overset{P}{\rightarrow}\bm{\beta}_0$ as $m \rightarrow \infty.$\\
(c) $\sqrt{m}(\hat{\bm{\beta}}-\bm{\beta}_0) \overset{\mathcal{D}}{\rightarrow} N(\bm{0},(-\mathbbm{E}\triangledown G(X_1,\bm{\beta}_0))^{-1})$ as $m \rightarrow \infty$.

\subsection{Proof of Lemma \ref{proposition: CUSUM approximation}}
\label{app: proof of CUSUM approximation}

From Assumption \ref{assumption: weight function}, $\underset{t \rightarrow 0}{\lim} t^{\gamma} \rho(t)<\infty$ implies $\rho(t)= \rho(k/m) = O((k/m)^{-\gamma})$ as $t = \frac{k}{m}\rightarrow 0$. Then $\omega(m,k) = m^{-\frac{1}{2}}\rho(k/m) = O(k^{-\gamma}m^{\gamma-\frac{1}{2}})$ as $\frac{k}{m}\rightarrow 0$. Similarly, as $t = \frac{k}{m}\rightarrow \infty$, $\omega(m,k) = O(m^{\frac{1}{2}}k^{-1}).$ 

Then there always exist a constant $C>0$ so that
\begin{equation}
\label{equation: weight function ineq}
\omega(m,k) \leq
\begin{cases}
C m^{\gamma-\frac{1}{2}}k^{-\gamma}, &\text{ when } k\leq m, \\
C m^{\frac{1}{2}}k^{-1}, &\text{ when } k>m.
\end{cases}    
\end{equation}

From (\ref{eq: estimatebeta}), we know that 
$$
\sum_{t=1}^m G(X_t,\hat{\bm{\beta}}) = 0,
$$
then the norm component of the lemma can be rewritten as
\small
$$
\begin{aligned}
&(\sum_{t=m+1}^{m+k} G(X_t,\hat{\bm{\beta}})-\sum_{t=m+1}^{m+k}G(X_t,\bm{\beta}_0))-(\frac{k}{m}\sum_{t=1}^m G(X_t,\hat{\bm{\beta}}) -\frac{k}{m}\sum_{t=1}^m G(X_t,\bm{\beta}_0)) \\
:=&D_1(m,k)-D_2(m,k).
\end{aligned}
$$
\normalsize
For the $j^{\text{th}}$ element in $G(X_t,\cdot)$, by Taylor expansion, the following equation holds
\small
\begin{equation}
\label{equation: taylor Hj}
\begin{aligned}
    &\sum_{t=m+1}^{m+k} G_j(X_t,\hat{\bm{\beta}})-\sum_{t=m+1}^{m+k} G_j(X_t,\bm{\beta}_0)\\ = &\sum_{t=m+1}^{m+k}(\triangledown G_j(X_t,\bm{\beta}_0)^T (\hat{\bm{\beta}}-\bm{\beta}_0)+ \frac{1}{2} (\hat{\bm{\beta}}-\bm{\beta}_0)^T \triangledown^2 G_j(X_t,\bm{\beta}^*)(\hat{\bm{\beta}}-\bm{\beta}_0)), \: j \in \{1,2, ..., l+2\},
\end{aligned}
\end{equation}
\normalsize
where $\bm{\beta}^*$ is between $\bm{\beta}_0$ and $\hat{\bm{\beta}}$.

The expression of the $j^{\text{th}}$ element of $G(X_t, \bm{\beta}^*)$ can be obtained from (\ref{eq: closed-form expression of G}).
$G_j(X_t, \bm{\beta}^*) = \mathbbm{Z}_{t-1,j}(X_t-n\pi_t(\bm{\beta}^*)).$ We can notice that the parameter $\bm{\beta}^*$ only exists in $\pi_t(\bm{\beta}^*)$, and $\pi_t(\bm{\beta}^*)$, determined by $\mathbbm{Z}_{t-1,j}$ and $\bm{\beta}^*$ through (\ref{equation: model definition1}), is a twice continuously differentiable function in terms of $\bm{\beta}^*$. It, together with the fact that every element in $\mathbbm{Z}_{t-1}$ is defined on a known and bounded space, implies that each element of the second derivative of $G_j(X_t,\bm{\beta}^*)$, $\triangledown^2 G_j(X_t,\bm{\beta}^*)$, is bounded. The following equation holds in a convex neighborhood of $\bm{\beta}_0$, $U_{\bm{\beta}_0} \subset \bm{B}$,
\begin{equation*}
\underset{\bm{\beta}^*\in U_{\bm{\beta}_0}}{\sup} ||\triangledown^2 G_j(X_t,\bm{\beta}^*)||_{\infty} =O(1), \text{ for } j \in \{1,2, ..., l+2\},
\end{equation*}
where $||.||_{\infty}$ is the maximum of the absolute values of all the elements of a matrix.
It follows that
\begin{equation}
\label{equation: sup Hj}
    \underset{k\geq 1}{\sup} \underset{\bm{\beta}^*\in U_{\bm{\beta}_0}}{\sup}\frac{1}{k}\sum_{t=m+1}^{m+k}||\triangledown^2 G_j(X_t,\bm{\beta}^*)||_{\infty} = O(1).
\end{equation}
Furthermore, by Birkhoff's Ergodic Theorem, the following equation holds 
\begin{equation}
\label{equation: Birkhoff Hj}
\frac{1}{k}\sum_{t=m+1}^{m+k} \triangledown G(X_t,\bm{\beta}_0)^T- \mathbbm{E}\triangledown G(X_1,\bm{\beta}_0)^T = o_p(1).
\end{equation}
Following the asymptotic normality of $\hat{\bm{\beta}}$ proven in Theorem \ref{prop: consistency&CLT},
\begin{equation*}
    \hat{\bm{\beta}}-\bm{\beta}_0 = O_p(\frac{1}{\sqrt{m}}) \text{ as } m \rightarrow \infty.
\end{equation*}
Then from (\ref{equation: weight function ineq})-(\ref{equation: Birkhoff Hj}), 
\begin{equation*}
\label{eq: D1,jth element}
\begin{aligned}
&\underset{k \geq 1}{\sup} \; \omega(m,k) |D_{1,j}(m,k)-k\mathbbm{E}\triangledown G_j(X_1,\bm{\beta}_0)^T(\hat{\bm{\beta}}-\bm{\beta}_0)| \\
= & \underset{k \geq 1}{\sup} \; \omega(m,k) |\sum_{t=m+1}^{m+k} G_j(X_t,\hat{\bm{\beta}})-\sum_{t=m+1}^{m+k}G_j(X_t,\bm{\beta}_0)-k\mathbbm{E}\triangledown G_j(X_1,\bm{\beta}_0)^T(\hat{\bm{\beta}}-\bm{\beta}_0)|\\
= & \underset{k \geq 1}{\sup} \; \omega(m,k) |\sum_{t=m+1}^{m+k}(\triangledown G_j(X_t,\bm{\beta}_0)^T (\hat{\bm{\beta}}-\bm{\beta}_0)+ \frac{1}{2} (\hat{\bm{\beta}}-\bm{\beta}_0)^T \triangledown^2 G_j(X_t,\bm{\beta}^*)(\hat{\bm{\beta}}-\bm{\beta}_0))-k\mathbbm{E}\triangledown G_j(X_1,\bm{\beta}_0)^T(\hat{\bm{\beta}}-\bm{\beta}_0)| \\
= &\underset{k \geq 1}{\sup} \; \omega(m,k) |\sum_{t=m+1}^{m+k}\triangledown G_j(X_t,\bm{\beta}_0)^T (\hat{\bm{\beta}}-\bm{\beta}_0)-k\mathbbm{E}\triangledown G_j(X_1,\bm{\beta}_0)^T(\hat{\bm{\beta}}-\bm{\beta}_0) \\
& \quad \quad \quad \quad \quad + \sum_{t=m+1}^{m+k}\frac{1}{2} (\hat{\bm{\beta}}-\bm{\beta}_0)^T \triangledown^2 G_j(X_t,\bm{\beta}^*)(\hat{\bm{\beta}}-\bm{\beta}_0)|\\
= &\underset{k \geq 1}{\sup} \; \omega(m,k)(o_p(\frac{k}{\sqrt{m}})+O_p(\frac{k}{m})) \\
=&\underset{k\leq \sqrt{m}}{\sup} \; O_p(k^{1-\gamma}m^{\gamma-\frac{3}{2}}) + \underset{\sqrt{m}<k\leq m}{\sup}o_p(\frac{k}{m})^{1-\gamma} + \underset{k>m}{\sup} \; O_p(\frac{1}{\sqrt{m}}) \\
= & o_p(1), \text{ for } j \in \{1,2,...,l+2\}.  
\end{aligned}
\end{equation*}
Therefore, 
  \begin{equation}
    \label{eq: D1}
    \underset{k \geq 1}{\sup} \; \omega(m,k) ||D_1(m,k)-k\mathbbm{E}\triangledown G(X_1,\bm{\beta}_0)^T(\hat{\bm{\beta}}-\bm{\beta}_0)|| = o_p(1). 
\end{equation}
Analogously, 
\begin{equation}
\label{eq: D2}
    \underset{k\geq 1}{\sup} \; \omega(m,k)||D_2(m,k)-k\mathbbm{E}\triangledown G(X_1,\bm{\beta}_0)^T(\hat{\bm{\beta}}-\bm{\beta}_0)|| = o_p(1).
\end{equation}
Considering (\ref{eq: D1}) and (\ref{eq: D2}), 
\small
$$
\begin{aligned}
&\underset{k \geq 1}{\sup} \; \omega(m,k) ||D_1(m,k)-D_2(m,k)|| \\
 = & \underset{k \geq 1}{\sup} \; \omega(m,k) ||(D_1(m,k)-k\mathbbm{E}\triangledown G(X_1,\bm{\beta}_0)^T(\hat{\bm{\beta}}-\bm{\beta}_0)) \\
 & -(D_2(m,k)-k\mathbbm{E}\triangledown G(X_1,\bm{\beta}_0)^T(\hat{\bm{\beta}}-\bm{\beta}_0))|| \\
 \leq & \underset{k \geq 1}{\sup} \; \omega(m,k) ||D_1(m,k)-k\mathbbm{E}\triangledown G(X_1,\bm{\beta}_0)^T(\hat{\bm{\beta}}-\bm{\beta}_0)|| \\
  &+ \underset{k\geq 1}{\sup} \; \omega(m,k)||D_2(m,k)-k\mathbbm{E}\triangledown G(X_1,\bm{\beta}_0)^T(\hat{\bm{\beta}}-\bm{\beta}_0)|| \\
  =&o_p(1).
\end{aligned}
$$
\normalsize

\subsection{Proof of Proposition \ref{proposition: wiener processes}}
\label{app: proof of wiener process}
We use the functional central limit theorem (Theorem 1 in \cite{oodaira1972functional}) to prove the proposition. According to Theorem 1 in \cite{oodaira1972functional}, if the process $\{G(X_t, \bm{\beta}_0)\}$ is 
\begin{itemize}
    \item bounded,
    \item strictly stationary,
    \item $\mathbbm{E}(G(X_t, \bm{\beta}_0)) = \bm{0}$,
    \item a strong mixing process,
\end{itemize} 
and satisfies the two inequalities (details will be discussed later)
\begin{equation*}
    \sum_{n=1}^{\infty} \alpha_G(n)<\infty
\end{equation*}
and 
\begin{equation*}
\alpha_G(n) \leq \frac{M_G}{n \log n },
\end{equation*}
$\{G(X_t, \bm{\beta}_0)\}$ converges weakly to a Wiener process with covariance matrix 
\begin{equation*}
    \bm{\Sigma} = \mathbbm{E}\big[G(X_1,\bm{\beta}_0)G^T(X_1,\bm{\beta}_0)\big]+2\sum_{t=2}^{\infty}\mathbbm{E} \big [G(X_1,\bm{\beta}_0)G^T(X_t,\bm{\beta}_0) \big ] <\infty.
\end{equation*}
Given the expression of $G$((\ref{eq: closed-form expression of G})) and the fact that every element in $\mathbbm{Z}_{t-1}$ is bounded, it is easy to verify that $\{G(X_t, \bm{\beta}_0)\}$ is bounded for any $t$. During the process of deriving (\ref{eq: closed-form expression of G}) we have concluded that $G(X_t, \bm{\beta}_0)$ is strictly stationary.

Since $X_t|\mathcal{C}_{t}\sim \text{Bin}(n, \pi_t(\bm{\beta}_0))$\footnote{The definition of $ \mathcal{C}_{t}$  can be found in the process of deriving (\ref{eq: closed-form expression of G}).}, we can see that 
\begin{equation*}
\begin{aligned}
    \mathbbm{E}(G(X_t, \bm{\beta}_0)) 
    &= \mathbbm{E}\Big(\mathbbm{E}(\mathbbm{Z}_{t-1}(X_t-n \pi_t(\bm{\beta}_0))|\mathcal{C}_{t})\Big) \\
    &= \mathbbm{E}\Big(\mathbbm{Z}_{t-1}\mathbbm{E}(X_t-n \pi_t(\bm{\beta}_0)|\mathcal{C}_{t})\Big)\\
    &= \bm{0}.
\end{aligned}
\end{equation*}
% It also shows that $G(X_t, \bm{\beta}_0)$ is a martingale difference sequence as $G$ is determined on $\mathcal{F}_{t-1}$.

Then we show the process $G(X_t, \bm{\beta}_0)$ is strong mixing.
In Theorem \ref{theorem:stationary&ergodic}, we have proved that the process $\{X_t\}$ is strictly stationary and geometrically ergodic. Given that $\{\bm{W}_t\}$ is an i.i.d. sequence, it is naturally strictly stationary and geometrically ergodic. So $\mathbbm{Z}_{t-1} = (1, X_{t-1},\bm{W}_t)$ is strictly stationary and geometrically ergodic. Then by Theorem 3.7 of \cite{bradley2005basic} (on page 121), if a strictly stationary Markov chain defined on a general state space is geometrically ergodic, the process has $\beta(n) = O_p(e^{-\theta n}), \theta>0$. For simplicity, we don't further introduce $\beta(n)$ here. What's worth pointing out is that according to (1.11) of \cite{bradley2005basic},  
\begin{equation*}
\label{eq: mixing inequality}
    2\alpha(n) \leq \beta(n),
\end{equation*}
where $\alpha(n)$ is the strong mixing coefficient. That is, for $\{\mathbbm{Z}_{t}\}$, there exists a $\theta>0$ so that the strong mixing coefficient 
\begin{equation*}
\begin{aligned}
    \alpha(n) &= \underset{j \in \mathbbm{N}^+}{\sup}\;\alpha(\mathcal{F}_0^j,\mathcal{F}_{j+n}^\infty) = \underset{j \in \mathbbm{N}^+}{\sup} \; |P(A \cap B)-P(A)P(B)|, \text{ for } \forall A \in \mathcal{F}_0^j, B \in \mathcal{F}_{j+n}^\infty \\
    &= O_p(e^{-\theta n}) \text{ as }n \rightarrow \infty,
\end{aligned}
\end{equation*}
where $\mathcal{F}_J^L := \sigma(\mathbbm{Z}_t, J \leq t \leq L).$ 
As $G(X_t, \bm{\beta}_0) = \mathbbm{Z}_{t-1}(X_t-n\pi_t(\bm{\beta}_0)),$ $\mathcal{G}_J^L = \sigma(G(X_t, \bm{\beta}_0), J \leq t \leq L) \subset \mathcal{F}_{J-1}^{L}$
Denote the strong mixing coefficient of $G(X_t,\bm{\beta}_0)$ as $\alpha_G(n)$, then it's easy to conclude that 
\begin{equation*}
    \begin{aligned}
    \alpha_G(n) = \underset{j \in \mathbbm{N}^+}{\sup}\;\alpha_G(\mathcal{G}_0^j,\mathcal{G}_{j+n}^\infty)\leq \underset{j \in \mathbbm{N}^+}{\sup}\;\alpha_G(\mathcal{F}_0^j,\mathcal{F}_{j+n-1}^\infty) = O_p(e^{-\theta (n-1)}) = O_p(e^{-\theta n}).
    \end{aligned}
\end{equation*}
As $n \rightarrow \infty, \alpha_G(n) \rightarrow 0,$ which means that $G(X_t, \bm{\beta}_0)$ is a strong mixing process.
We can find a $n_0>0$ and the corresponding $M_0$ such that the inequality $\alpha_G(n)\leq M_0 e^{-\theta n}$ holds for $n\geq n_0$. Let $M_-:= \underset{n=1,2,...,n_0-1}{\max} \alpha_G(n) e^{\theta n}$ and let $M = \max (M_0,M_-)$, then 
\begin{equation}
\label{eq: alpha_G}
\alpha_G(n) \leq M e^{-\theta n}, \forall n \in \mathbbm{N}^+
\end{equation}
holds.
Based on (\ref{eq: alpha_G}), the inequality 
\begin{equation}
\label{eq: sum of alpha_G is bounded}
    \sum_{n=1}^{\infty} \alpha_G(n) \leq \sum_{n=1}^{\infty} Me^{-\theta n}<\infty
\end{equation}
holds.

From (\ref{eq: alpha_G}), we can see that it's not hard to find a $M_G>0$ such that
\begin{equation}
\label{eq: alpha_G is bounded}
\alpha_G(n) \leq \frac{M_G}{n \log n }.
\end{equation}
First we define the function $f(n) = M e^{-\theta n}n \log n, n \in \mathbbm{N}^+$. Then $[\log f(n)]' = -\theta+\frac{1}{n}+\frac{1}{n \log n}$ is a decreasing function and there always exists a $n_{\theta}$ such that for $n >n_{\theta}$, $[\log f(n)]'<0$, which means we can always find $m = \underset{n \in \mathbbm{N}^+}{\max}f(n)$. Therefore, as long as we choose $M_G \geq m$, (\ref{eq: alpha_G is bounded}) holds.

According to Theorem 1 in \cite{oodaira1972functional}, (\ref{eq: sum of alpha_G is bounded}) and (\ref{eq: alpha_G is bounded}) imply that

\begin{equation*}
    \left\{\frac{1}{\sqrt{m}}\sum_{t=1}^{\floor{ms}} G(X_t,\bm{\beta}_0)\right\}\overset{\mathcal{D}}{\rightarrow}W(s), \text{ as }m \rightarrow \infty,
\end{equation*}
where $W(s)$ is a Wiener process with covariance $\bm{\Sigma}$
\begin{equation*}
    \bm{\Sigma} = \mathbbm{E}\big[G(X_1,\bm{\beta}_0)G^T(X_1,\bm{\beta}_0)\big]+2\sum_{t=2}^{\infty}\mathbbm{E} \big [G(X_1,\bm{\beta}_0)G^T(X_t,\bm{\beta}_0) \big ] <\infty.
\end{equation*}
For any $t>1$,
\begin{equation*}
\begin{aligned}
\label{eq: G MDS}
    \mathbbm{E} [G(X_1,\bm{\beta}_0)G^T(X_t,\bm{\beta}_0) ] 
    &= \mathbbm{E}\Big[\mathbbm{E}[G(X_1,\bm{\beta}_0)G^T(X_t,\bm{\beta}_0)|\mathcal{C}_{t}] \Big ] \\
    &=  \mathbbm{E}\Big[G(X_1,\bm{\beta}_0)\mathbbm{Z}_{t-1}\mathbbm{E}[(X_t-n\pi_t(\bm{\beta}_0))|\mathcal{C}_t]\Big ] \\
    &= \bm{0}.
\end{aligned}
\end{equation*}
% Also, based on (\ref{eq: sample information matrix}), 
% \begin{equation}
% \label{eq: fisher information of G}
% \mathbbm{E}\big[G(X_1,\bm{\beta}_0)G^T(X_1,\bm{\beta}_0)\big] = -\mathbbm{E}\triangledown G(X_1,\bm{\beta}_0).
% \end{equation}
The covariance matrix can be simplified as 
\begin{equation*}
    \bm{\Sigma} = \mathbbm{E}\big[G(X_1,\bm{\beta}_0)G^T(X_1,\bm{\beta}_0)\big] .
    % = -\mathbbm{E}\triangledown G(X_1,\bm{\beta}_0).
\end{equation*}

\subsection{Proof of Theorem \ref{theorem: null}}
\label{app: proof of the null}
We continue to use the notation introduced in \cite{kirch2015use} that $Z^TAZ:=||Z||^2_A$. \\
By using Lemma \ref{proposition: CUSUM approximation}, the test statistic
\small
\begin{equation}
\begin{aligned}
\label{equation: close-end test statistics 1}
        &\underset{1\leq k \leq Nm}{\sup} \omega(m,k)^2 ||S(m,k)||^2_A \\
        =&\underset{1\leq k \leq Nm}{\sup} \omega(m,k)^2||\sum_{t=m+1}^{m+k}G(X_t,\bm{\beta}_0)-\frac{k}{m}\sum_{t=1}^m G(X_t,\bm{\beta}_0)||^2_A+o_p(1) \\
        =&\underset{0<k/m \leq N}{\sup}(\frac{1}{\sqrt{m}}\rho(k/m))^2 ||\sum_{t=m+1}^{m+k}G(X_t,\bm{\beta}_0)-\frac{k}{m}\sum_{t=1}^m G(X_t,\bm{\beta}_0)||^2_A+o_p(1)\\
        =&\underset{0<s \leq N}{\sup}\rho(s)^2 ||\frac{1}{\sqrt{m}}\sum_{t=m+1}^{m+k}G(X_t,\bm{\beta}_0)-\frac{1}{\sqrt{m}}\frac{k}{m}\sum_{t=1}^m G(X_t,\bm{\beta}_0)||^2_A+o_p(1).\\
\end{aligned}
\end{equation}
\normalsize
From Proposition \ref{proposition: wiener processes} we know that, with $k = \lfloor ms \rfloor$, as $m \rightarrow \infty$,
\begin{equation}
\label{equation: close-end G1}
    \frac{1}{\sqrt{m}}\sum_{t=m+1}^{m+k}G(X_t,\bm{\beta}_0) = \frac{1}{\sqrt{m}}(\sum_{t=1}^{m+k}G(X_t,\bm{\beta}_0)-\sum_{t=1}^m G(X_t,\bm{\beta}_0)) \overset{\mathcal{D}}{\rightarrow} W_1(k/m):= W_1(s),
\end{equation}
and
\begin{equation}
\label{equation: close-end G2} 
    \frac{1}{\sqrt{m}}\frac{k}{m} \sum_{t=1}^m G(X_t,\bm{\beta}_0) \overset{\mathcal{D}}{\rightarrow} s W_2(1).
\end{equation}
$W_1(.)$ and $W_2(.)$ are Wiener processes defined in Proposition \ref{proposition: wiener processes}. 

The continuity of $\rho^2(s)||W_1(s)-sW_2(1)||^2_A$ on $s\in [0,N]$, together with (\ref{equation: close-end test statistics 1}), (\ref{equation: close-end G1}) and (\ref{equation: close-end G2}) yield that
\begin{equation*}
\begin{aligned}
    \underset{m \rightarrow \infty}{\lim}&P\left(\underset{1\leq k \leq Nm}{\sup} \omega(m,k)^2S(m,k)^T\bm{A}S(m,k) \leq c    \right) \\
     = &P \left( \underset{0<s\leq N}{\sup} \rho^2(s)(W_1(s)-sW_2(1))^T\bm{A}(W_1(s)-sW_2(1))\leq c \right)
\end{aligned}
\end{equation*}
for a given threshold $c$.

\subsection{Proof of Theorem \ref{theorem: power analysis}}
\label{app: proof of power analysis}
We denote the change point as $k^*$, and use $\Tilde{k} = \lfloor m x_0 \rfloor$ to denote a data point after $k^*$. Then 
\begin{equation*}
\begin{aligned}
        \bm{S}(m,\Tilde{k}) &= \sum_{t=m+1}^{m+k^*} G(X_t,\hat{\bm{\beta}})+ \sum_{t=m+k^*+1}^{m+\Tilde{k}} G(X_t^*,\hat{\bm{\beta}})\\
                            &=: \bm{S}_{H_0}(m,k^*)+\bm{S}_{H_a}(m+k^*,\Tilde{k}-k^*)
\end{aligned}
\end{equation*}
Under Assumption \ref{assumption: power analysis}(a), as $m \rightarrow \infty$, $k^* \rightarrow \infty$, due to ergodicity of $\{X_t\}$ and consistency of $\hat{\bm{\beta}}$,
\begin{equation*}
    \frac{1}{k^*} \bm{S}_{H_0}(m,k^*) \overset{P}{\rightarrow}0,
\end{equation*}
and
\begin{equation}
\label{eq: power proof 1}
    \frac{1}{m} \bm{S}_{H_0}(m,k^*) \overset{P}{\rightarrow}0.
\end{equation}
After the change point $k^*$, under Assumption \ref{assumption: power analysis}(c), 
\begin{equation}
\label{eq: power proof 2}
    \frac{1}{m}\bm{S}_{H_a}(m+k^*,\Tilde{k}-k^*) = \frac{x_0-\nu}{(x_0-\nu)m} \sum_{t = m+k^*+1}^{m+\Tilde{k}}G(X_t^*,\hat{\bm{\beta}}) \overset{P}{\rightarrow} (x_0-\nu) \bm{E}_{G,H_a}.
\end{equation}
Combining (\ref{eq: power proof 1}) and (\ref{eq: power proof 2}), 
\begin{equation*}
    \begin{aligned}
    & \underset{1\leq k \leq Nm}{\sup}\omega^2(m,k) \bm{S}(m,k)^T\bm{A}\bm{S}(m,k) \\
    \geq & \frac{1}{m}\rho^2(x_0) (\bm{S}_{H_0}(m,k^*)+\bm{S}_{H_a}(m+k^*,\Tilde{k}-k^*))^T \bm{A} (\bm{S}_{H_0}(m,k^*)+\bm{S}_{H_a}(m+k^*,\Tilde{k}-k^*)) \\
    =& m \rho^2(x_0) (\frac{1}{m}\bm{S}_{H_0}(m,k^*)+\frac{1}{m}\bm{S}_{H_a}(m+k^*,\Tilde{k}-k^*))^T \bm{A} (\frac{1}{m}\bm{S}_{H_0}(m,k^*)+\frac{1}{m}\bm{S}_{H_a}(m+k^*,\Tilde{k}-k^*)) \\
     = & m\rho^2(x_0)(x_0-\nu)^2 (\bm{E}_{G,H_a}+o_p(1))^T \bm{A} (\bm{E}_{G,H_a}+o_p(1)) \\
     \rightarrow & \infty.
    \end{aligned}
\end{equation*}
Therefore, with any fixed threshold $c$,
\begin{equation*}
    \underset{m \rightarrow \infty}{\lim}P\left(\underset{1\leq k \leq Nm}{\sup} \omega(m,k)^2\bm{S}(m,k)^T\bm{A}\bm{S}(m,k) \geq c  |H_a  \right) = 1.
\end{equation*}

\bibliography{arxiv} 
\bibliographystyle{ieeetr}
\end{document}